\documentclass[preprint,12pt,times]{elsarticle}

\usepackage{geometry}
\usepackage{amssymb,amsmath,amsthm}
\usepackage{graphicx}
\usepackage{booktabs}
\usepackage{multirow}
\usepackage{array}
\usepackage{bm}
\usepackage{xcolor}
\usepackage{url}
\usepackage{microtype}
\usepackage{enumitem}

\usepackage{subcaption}
\usepackage[font=small,labelfont=bf,labelsep=period,
            justification=justified,singlelinecheck=false]{caption}
\usepackage{float}
\usepackage{placeins}
\usepackage{needspace}

\usepackage[hidelinks]{hyperref}


\newtheorem{proposition}{Proposition}
\newtheorem{lemma}{Lemma}
\theoremstyle{remark}
\newtheorem{remark}{Remark}
\newcommand{\nrms}[1]{\lVert #1\rVert_{\mathrm{rms}}}

\journal{Computer Physics Communications}

\begin{document}

\begin{frontmatter}

\title{CoPiT: A conditional Position-induced Transformer for aerodynamic field prediction on unstructured meshes}

\author[hkust]{Junfeng Chen\corref{cor1}}\ead{majfchen@ust.hk}
\author[polyu]{Yongchen Fan\corref{cor1}}
\ead{23040781r@connect.polyu.hk}
\author[polyu]{Zhonghua Qiao}
\author[sustech]{Kailiang Wu}

\cortext[cor1]{Corresponding authors}
\address[polyu]{Department of Applied Mathematics, The Hong Kong Polytechnic University, Hung Hom, Kowloon, Hong Kong, China}
\address[hkust]{Department of Mathematics, The Hong Kong University of Science and Technology, Clear Water Bay, Kowloon, Hong Kong, China}
\address[sustech]{Department of Mathematics and Shenzhen International Center for Mathematics, Southern University of Science and Technology, Shenzhen 518055, China}

\begin{abstract}
High-fidelity Reynolds-averaged Navier--Stokes (RANS) simulations remain a standard tool for aerodynamic analysis, yet each solve on an unstructured mesh is computationally expensive, making many-query tasks such as design exploration, optimization, and multi-condition evaluation costly. Neural operators provide mesh-flexible surrogates for mapping geometry and operating conditions to flow fields; however, in many existing architectures, global parameters such as the freestream state and design variables are concatenated with pointwise inputs, leaving their influence on intermediate features implicit. This paper proposes CoPiT, a \textbf{Co}nditional \textbf{P}osition-induced \textbf{T}ransformer that builds on the Position-induced Transformer (PiT) as its geometric backbone and injects a global condition vector through multi-stage feature-wise linear modulation (FiLM), thereby making parametric modulation explicit while preserving geometry-based attention weights. The model is evaluated on two-dimensional AirfRANS airfoil fields, including scarce-data and extrapolation regimes, and on three-dimensional BlendedNet++ blended-wing-body surfaces. On the AirfRANS full-data test set, CoPiT reduces surface and volume losses by approximately 54\% and 24\% relative to Transolver, respectively, and achieves the lowest overall normalized test MSE among the compared baselines on BlendedNet++. Matched ablations show that FiLM conditioning generally improves upon input concatenation, while the preferred placement is task-dependent. CoPiT thus provides a practical operator-learning surrogate for parametric aerodynamics on unstructured meshes.
\end{abstract}


\end{frontmatter}

\section{Introduction}
\label{sec:intro}
\begin{figure}[!t]
\centering
\includegraphics[width=0.9\linewidth,keepaspectratio]{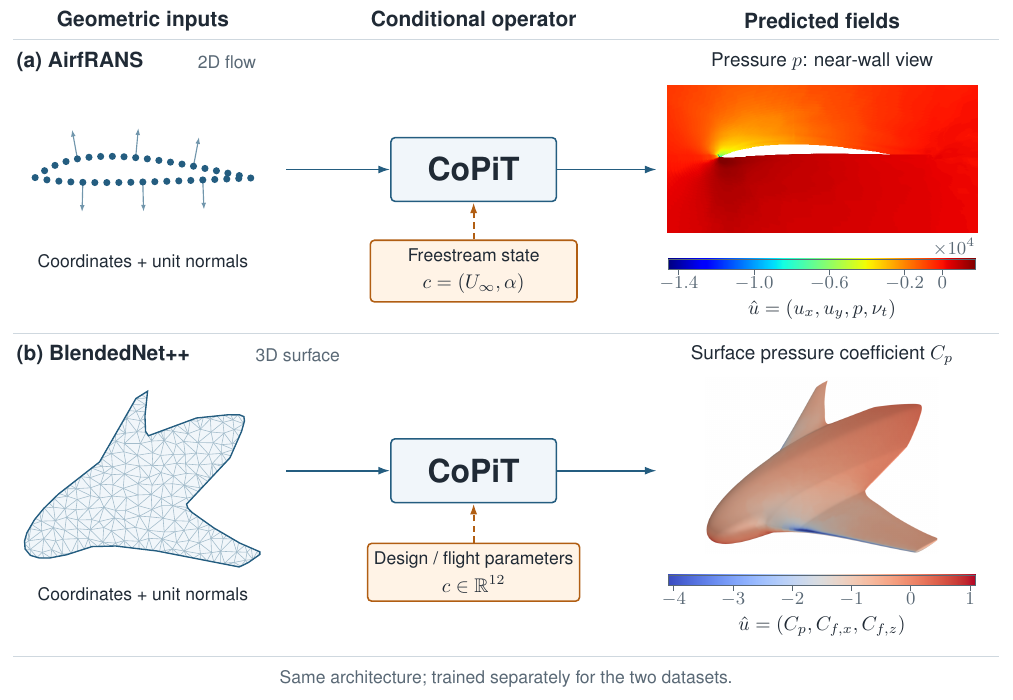}
\caption{Schematic of a surrogate-assisted aerodynamic design workflow. Offline CFD generates training data; a learned operator then provides approximate evaluations during design exploration.}
\label{fig:surrogate_pipeline}
\end{figure}

Parametric partial differential equations (PDEs) are central to fluid mechanics and aerospace applications~\cite{anderson2017fundamentals,vinuesa2022cfd_ml}. In particular, high-fidelity Reynolds-averaged Navier--Stokes (RANS) simulations yield velocity, pressure and wall-shear fields that are indispensable for aerodynamic analysis, yet each solve is computationally expensive. Many-query tasks---design exploration, optimization and multi-condition evaluation---therefore motivate surrogate models that approximate the map from geometry and freestream or design parameters to aerodynamic fields at a fraction of the CFD cost~\cite{brunton2020ml_fluid,karniadakis2021piml}. Figure~\ref{fig:surrogate_pipeline} illustrates this surrogate-assisted workflow.

Neural operators learn mappings between function spaces and, once trained, evaluate new inputs by a single forward pass~\cite{li2021fno,lu2021deeponet,kovachki2023neural}. Fourier Neural Operators (FNO)~\cite{li2021fno} and DeepONet~\cite{lu2021deeponet} are representative neural-operator architectures, while Geo-FNO and GINO extend operator learning to more general geometries~\cite{li2023geofno,li2022gino}. Graph networks and transformer-based solvers further address unstructured meshes~\cite{pfaff2021learning,wu2024transolver,luo2025transolverpp,chen2024pit}. These developments make operator learning a natural framework for aerodynamic field prediction, where the surrogate must remain accurate across varying meshes and parametric regimes.

Aerodynamic field prediction on irregular geometries presents two coupled challenges. First, the discrete domain is geometry-dependent (unstructured meshes or surface triangulations) rather than a fixed Cartesian grid. Second, the solution depends jointly on local geometry and global parameters such as freestream speed, angle of attack and design variables~\cite{bhatnagar2019cnn_aero,thuerey2020deepcfd,bonnet2022airfrans,sung2025blendednetpp}. These two dependencies motivate distinct architectural roles for geometry-based spatial aggregation and condition-dependent feature modulation. Geometry-aware architectures (MeshGraphNets~\cite{pfaff2021learning}, Transolver~\cite{wu2024transolver}, PiT~\cite{chen2024pit}) capture spatial interactions effectively, but when global parameters such as the freestream state and design variables are supplied only through input concatenation, their influence on intermediate representations remains implicit. Such one-shot fusion does not explicitly distinguish geometry-driven aggregation from parametric response modulation. Whether explicit conditioning at intermediate stages improves field and integrated-load prediction, and how its placement affects different metrics, therefore remain empirical questions.

Feature-wise Linear Modulation (FiLM) offers a lightweight alternative: a small network maps a parameter vector to channel-wise scale and shift coefficients that modulate intermediate features~\cite{perez2018film,dumoulin2018film_survey}. This provides a means of explicitly conditioning intermediate representations at multiple stages, rather than relying solely on input-level parameter injection.

We propose CoPiT, a \textbf{Co}nditional \textbf{P}osition-induced \textbf{T}ransformer for aerodynamic field prediction. CoPiT retains PiT's latent-mesh encode--process--decode architecture, in which position-induced attention weights are determined by pairwise spatial distances rather than feature values~\cite{chen2024pit}. It introduces multi-stage FiLM conditioning after instance normalization in every processor block and in the decoder, allowing global conditions to modulate intermediate features without altering the geometry-based attention weights. We evaluate CoPiT on AirfRANS~\cite{bonnet2022airfrans} (two-dimensional RANS airfoil fields; full-data, scarce-data, and extrapolation settings) and BlendedNet++~\cite{sung2025blendednetpp} (three-dimensional blended-wing-body surface fields). Dataset-specific encoder inputs and conditioning configurations are detailed in Section~\ref{sec:experiments}. We further analyze the properties of this conditioning mechanism, including its effect on position-attention and the dependence of predictions on global conditions.

The contributions of this work are as follows:
\begin{itemize}[leftmargin=1.4em,itemsep=0.2em,topsep=0.3em]
\item \textbf{Architecture.} We propose CoPiT, which combines PiT's geometry-based position-attention with multi-stage FiLM conditioning in the processor and decoder, enabling explicit modulation of intermediate features by global parameters.

\item \textbf{Analysis.} We show that pointwise FiLM does not modify position-attention weights and preserves convergence of an already convergent attention output under bounded modulation. Under FiLM-only conditioning, the encoder features and all attention weights are independent of the condition for fixed geometric inputs, and the prediction is Lipschitz continuous in the condition, with an explicit bound depending on the FiLM generators and backbone components; see Propositions~\ref{prop:separation} and~\ref{prop:lipschitz}.

\item \textbf{Evaluation.} We evaluate CoPiT on two-dimensional AirfRANS and three-dimensional BlendedNet++, including scarce-data and extrapolation settings on AirfRANS. On the AirfRANS full-data test set, CoPiT reduces surface and volume losses by approximately 54\% and 24\%, respectively, relative to Transolver; on BlendedNet++, it achieves the lowest overall normalized test MSE among the reimplemented baselines. Matched conditioning ablations show that FiLM generally outperforms input concatenation, while the preferred FiLM placement depends on the evaluation metric.
\end{itemize}

The remainder of the paper is organized as follows. Section~\ref{sec:related} reviews related work. Section~\ref{sec:prelim} states the learning problem and reviews position-induced attention. Section~\ref{sec:method} introduces CoPiT. Section~\ref{sec:experiments} reports numerical experiments. Section~\ref{sec:conclusion} concludes.

\section{Related work}
\label{sec:related}

Neural operators approximate solution maps of parametric PDEs from data~\cite{kovachki2023neural}. DeepONet represents the operator as a branch--trunk product~\cite{lu2021deeponet}; Fourier Neural Operators (FNO) mix features in the frequency domain on regular grids~\cite{li2021fno}. Geo-FNO~\cite{li2023geofno} and GINO~\cite{li2022gino} extend related ideas to deformed or unstructured geometries. Graph networks operate directly on mesh connectivity~\cite{pfaff2021learning,chen2021graph}. Transformer-based neural operators~\cite{cao2021choose,li2023transformer,hao2023gnot} adopt different attention mechanisms to handle general geometries and heterogeneous inputs. Transolver and Transolver++ use physics-aware attention over learned representations~\cite{wu2024transolver,luo2025transolverpp}, whereas the Position-induced Transformer (PiT) constructs attention weights from pairwise spatial distances~\cite{chen2024pit}.

Data-driven aerodynamic surrogates include convolutional predictors of airfoil fields~\cite{bhatnagar2019cnn_aero,thuerey2020deepcfd}. Public benchmarks such as AirfRANS~\cite{bonnet2022airfrans} and BlendedNet++~\cite{sung2025blendednetpp} provide two-dimensional airfoil flow fields and three-dimensional blended-wing-body surface fields, respectively. These benchmarks provide complementary settings for assessing the role of explicit conditioning in geometry-aware aerodynamic surrogates.

Global parameters can be incorporated by concatenating a broadcast condition vector with pointwise features. Feature-wise Linear Modulation (FiLM) instead maps a condition vector to channel-wise scale and shift coefficients that modulate intermediate features~\cite{perez2018film,dumoulin2018film_survey}. In aerodynamic surrogate modeling, BlendedNet~\cite{sung2025blendednet} uses FiLM for pointwise surface-field prediction, while the BlendedNet++ benchmark includes a FiLMNet model without position-induced attention~\cite{sung2025blendednetpp}. CoPiT combines PiT's geometry-based position-induced attention with multi-stage FiLM conditioning, allowing global parameters to modulate intermediate features without altering the distance-based attention weights.

\section{Preliminaries}
\label{sec:prelim}
\subsection{Operator learning}
\label{sec:op_learn}
Consider a parametric PDE of the form
\begin{equation}
  \mathcal{L}_{a}u = f,
  \label{eq:pde}
\end{equation}
where
\begin{equation}
  a \in \mathcal{A}(\Omega_{a};\mathbb{R}^{d_{a}}),
  \qquad
  u \in \mathcal{U}(\Omega_{u};\mathbb{R}^{d_{u}}),
  \qquad
  f \in \mathcal{F}.
  \label{eq:spaces}
\end{equation}
Here $\Omega_{a},\Omega_{u}\subset\mathbb{R}^{d}$ are spatial sets (not necessarily identical) that may represent volumetric domains or surfaces; $\mathcal{A}$ and $\mathcal{U}$ are Banach spaces of functions over $\Omega_{a}$ and $\Omega_{u}$, respectively, and $\mathcal{L}_{a}:\mathcal{U}\to\mathcal{F}$ is a differential operator depending on the parameter function $a$. Assuming suitable boundary conditions and well-posedness, the associated solution operator is
\begin{equation}
  \Phi : \mathcal{A} \to \mathcal{U},
  \qquad
  u = \Phi(a).
  \label{eq:solution_op}
\end{equation}
Operator learning aims to construct a parameterized surrogate $\Phi_{\theta}$ such that $\Phi_{\theta}(a)\approx\Phi(a)$ for previously unseen inputs~\cite{kovachki2023neural}.

In applications, $a$ and $u$ are observed only through samples on finite meshes
\begin{equation}
  X_a^{(j)}
=\left\{x_i^{(j)}\right\}_{i=1}^{N_a^{(j)}},
\qquad
X_u^{(j)}
=\left\{\hat x_i^{(j)}\right\}_{i=1}^{N_u^{(j)}}.
  \label{eq:meshes}
\end{equation}
Given training pairs $\{(a^{j},u^{j})\}_{j=1}^{J}$, with $a^{j}$ and $u^{j}$ sampled on the corresponding input and output point sets \(X_a^{(j)}\) and \(X_u^{(j)}\), respectively, the parameters $\theta$ can be learned by minimizing an empirical loss, for example,
\begin{equation}
  \min_{\theta}\;
  \frac{1}{J}
  \sum_{j=1}^{J}
  \bigl\|
    u^{j}
    -
    \Phi_{\theta}(a^{j})
  \bigr\|^{2},
  \label{eq:emp_risk}
\end{equation}
where the norm is evaluated on the corresponding output point set (or on a subsample of it). Once trained, \(\Phi_\theta\) is designed to accept inputs sampled on different meshes and produce predictions at specified query locations in \(\Omega_u\).

In the present work the parameter function is induced by a geometry $g$ and a global parameter vector $\mathbf{c}\in\mathbb{R}^{d_{c}}$ collecting freestream and design quantities. We write $a=a_{g,\mathbf{c}}$. For varying geometries, the input and output domains, and hence the corresponding function spaces, may depend on $g$; the operator notation above is understood accordingly. The output fields considered later are
\begin{align}
  u_{g,\mathbf{c}}(x)
  &=
  \bigl(u_{x}(x),\, u_{y}(x),\, p(x),\, \nu_{t}(x)\bigr)
  &&
  \text{(AirfRANS)},
  \label{eq:AirfRANS_target}\\
  u_{g,\mathbf{c}}(x)
  &=
  \bigl(C_{p}(x),\, C_{f,x}(x),\, C_{f,z}(x)\bigr)
  &&
  \text{(BlendedNet++)},
  \label{eq:bwb_target}
\end{align}
and the learning task is to approximate $u_{g,\mathbf{c}}$ by $\Phi_{\theta}(a_{g,\mathbf{c}})$ under the corresponding data protocol.

\subsection{Position-induced attention}
\label{sec:posatt}
A standard building block of Transformer architectures is content-based self-attention~\cite{vaswani2017attention}. Query, key and value projections of a feature matrix are combined through a softmax over pairwise feature affinities, so the weights depend on the current field values, must be recomputed for every input, and do not directly encode the geometry of the underlying mesh. For PDE surrogates, however, the relative positions of sampling points provide a useful inductive bias for feature aggregation.

Position-induced attention incorporates such geometric structure in an explicit and efficient way~\cite{chen2024pit}. Let $\mathbf{U}\in\mathbb{R}^{N\times d_{\ell}}$ collect the values of a feature field on a mesh $X=\{x_{i}\}_{i=1}^{N}$, and define the pairwise squared-distance matrix
\begin{equation}
  S_{ij}(X)
  =
  \|x_{i}-x_{j}\|_{2}^{2},
  \qquad
  i,j=1,\ldots,N.
  \label{eq:dist_mat}
\end{equation}
With a nonnegative learnable kernel scale $\lambda\ge 0$ and a value projection matrix $W^{V}$, the basic position-attention operator is
\begin{equation}
  \operatorname{PosAtt}(X,\mathbf{U})
  =
  \operatorname{Softmax}(-\lambda \mathbf{S}(X))\,
  \mathbf{U} W^{V}.
  \label{eq:posatt}
\end{equation}
Each row of the output is therefore a distance-weighted average of value-transformed features: the kernel depends on mesh geometry rather than on the current feature content. Multi-head variants use a kernel scale $\lambda_{h}=\varphi(\tilde\lambda_{h})\ge 0$ on head $h$, where $\varphi$ is a nonnegative reparameterization of an unconstrained parameter $\tilde\lambda_{h}$, and the corresponding operator is written $\operatorname{PosAtt}_{h}$. When the query mesh $X$ and the key mesh $Y$ differ, the rectangular matrix $S_{ij}(X,Y)=\|x_{i}-y_{j}\|_{2}^{2}$ defines cross position-attention $\operatorname{PosAtt}_{h}(X,Y,\mathbf{U})$, with the rows of $\mathbf{U}$ on $Y$. Self position-attention is the special case $\operatorname{PosAtt}_{h}(X,\mathbf{U})=\operatorname{PosAtt}_{h}(X,X,\mathbf{U})$. A locality parameter $\tau\in(0,1]$ may further restrict attention to a nearest-neighbor fraction of keys (quantile masking), while $\tau=1$ recovers a global kernel.

A central property of~\eqref{eq:posatt} is discretization consistency: under the sampling and boundedness assumptions of Theorem~2.1 in~\cite{chen2024pit}, its discrete outputs converge in probability to those of an integral operator with a normalized Gaussian-type kernel as the number of sampling points increases. This property makes position-attention an attractive geometric primitive for mesh-based aerodynamic surrogates.

In the PiT architecture~\cite{chen2024pit}, position-attention is organized in an encoder--processor--decoder form with a latent mesh of moderate size. Cross position-attention maps the input mesh to the latent mesh and, later, the latent representation to the output mesh, while a stack of self position-attention blocks refines features on the latent mesh. The next section builds CoPiT on this geometric organization.

\section{CoPiT: conditional position-induced Transformer}
\label{sec:method}
\subsection{Overview}
\label{sec:overview}
\begin{figure}[t]
\centering
\includegraphics[width=\linewidth,keepaspectratio]{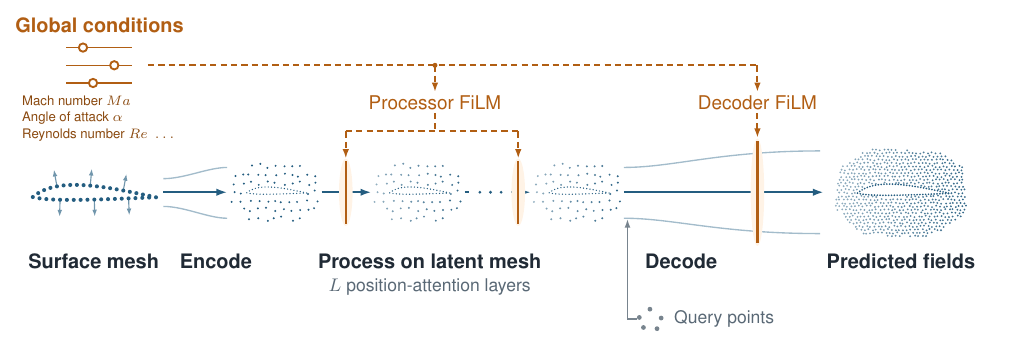}
\caption{Overview of CoPiT architecture. Surface features are encoded onto a latent mesh, then refined by \(L\) position-attention layers, and decoded at query points. Global conditions modulate processor and decoder features through FiLM.}
\label{fig:copit_arch}
\end{figure}

CoPiT builds on the latent-mesh encoder--processor--decoder architecture of PiT, incorporating global conditions through FiLM at intermediate stages (Fig.~\ref{fig:copit_arch}). Given an input mesh with pointwise features, a target mesh and a condition vector $\mathbf{c}$, the network (i)~encodes the input onto a latent mesh, (ii)~refines latent features with $L$ processor blocks containing FiLM modulation, and (iii)~decodes to the target mesh with a final FiLM stage before the readout. The same architectural skeleton is used for AirfRANS and BlendedNet++; dataset-specific encoder inputs, the components of $\mathbf{c}$ and auxiliary decoder channels are detailed in Section~\ref{sec:setup}.

\subsection{Multi-stage FiLM conditioning}
\label{sec:film}
Let $q$ denote the number of channels being modulated, and let $\mathbf{h}\in\mathbb{R}^{q}$ be the feature at one mesh or latent point after instance normalization~\cite{ulyanov2016instance}. The normalization is computed channel-wise across the points of the feature matrix, as in~\eqref{eq:instnorm}, rather than on an isolated feature vector, and $\mathbf{h}$ is the row at that point. A FiLM generator $g_{\theta}:\mathbb{R}^{d_{c}}\to\mathbb{R}^{2q}$ maps $\mathbf{c}$ to one scale--shift pair $(\gamma(\mathbf{c}),\beta(\mathbf{c}))$ and applies
\begin{equation}
  \tilde{\mathbf{h}}
  =
  \gamma(\mathbf{c})\odot\mathbf{h}+\beta(\mathbf{c}).
  \label{eq:film}
\end{equation}
\begin{figure}[t]
\centering
\includegraphics[width=\linewidth,keepaspectratio]{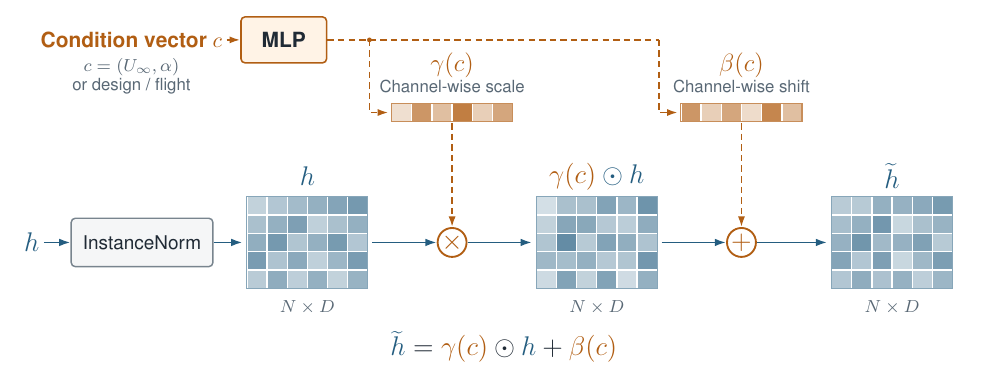}
\caption{Feature-wise Linear Modulation (FiLM)~\cite{perez2018film} in CoPiT. A global condition vector $\mathbf{c}$ is mapped to channel-wise scale $\gamma(\mathbf{c})$ and shift $\beta(\mathbf{c})$, which modulate instance-normalized features. The width $D$ printed on the feature matrices is the number of channels being modulated, denoted $q$ in the text.}
\label{fig:film}
\end{figure}
Figure~\ref{fig:film} illustrates the mechanism~\cite{perez2018film}. CoPiT uses two generators:
\begin{itemize}
\item \emph{Processor FiLM}: outputs $\{(\gamma_{\ell},\beta_{\ell})\}_{\ell=1}^{L}$, with each vector in $\mathbb{R}^{D_{\mathrm{p}}}$;
\item \emph{Decoder FiLM}: outputs $(\gamma_{\mathrm{dec}},\beta_{\mathrm{dec}})$, with each vector in $\mathbb{R}^{D_{\mathrm{d}}}$.
\end{itemize}
The widths $D_{\mathrm{p}}$ and $D_{\mathrm{d}}$ are fixed in Section~\ref{sec:blocks}. The implemented processor generator is $g_{\mathrm{proc}}:\mathbb{R}^{d_{c}}\to\mathbb{R}^{2L\,D_{\mathrm{p}}}$, split into the $L$ pairs above, and the decoder generator is $g_{\mathrm{dec}}:\mathbb{R}^{d_{c}}\to\mathbb{R}^{2D_{\mathrm{d}}}$.
Each generator is a small MLP (two hidden layers of width $64$ in our experiments) trained end-to-end. Thus, the condition modulates features throughout the processor and again in the decoder, rather than only at the input.

The affine transformation in~\eqref{eq:film} has the following basic properties; the approximation statement in (iii) follows from universal approximation for multilayer perceptrons~\cite{hornik1989mlp}.
\begin{proposition}[Basic properties of FiLM]
\label{prop:film}
Let $\tilde{\mathbf{h}}=\gamma(\mathbf{c})\odot\mathbf{h}+\beta(\mathbf{c})$. Then:
\begin{enumerate}
\item[(i)] $\gamma\equiv\mathbf{1}$ and $\beta\equiv\mathbf{0}$ recover the unmodulated feature;
\item[(ii)] $\gamma_{k}=\beta_{k}=0$ suppresses channel $k$;
\item[(iii)] on any compact set of conditions, continuous maps $\mathbf{c}\mapsto(\gamma(\mathbf{c}),\beta(\mathbf{c}))$ can be approximated uniformly by a feed-forward generator of sufficient width~\cite{hornik1989mlp}.
\end{enumerate}
\end{proposition}

\subsection{Processor and decoder with FiLM}
\label{sec:blocks}
We now specify how FiLM enters the PiT processor and decoder. Denote the latent feature at one point after encoding by $\mathbf{f}_{\mathrm{ltt}}^{(0)}\in\mathbb{R}^{D}$, where $D$ is the hidden width, and let $\mathbf{F}^{(\ell)}\in\mathbb{R}^{N_{\mathrm{ltt}}\times D}$ be the matrix whose rows are these features on $X_{\mathrm{ltt}}$. In the reported implementation every value projection in~\eqref{eq:posatt} is the identity, and channel mixing is performed by the multilayer perceptron in~\eqref{eq:proc4}. For block $\ell$ and head $h$,
\begin{equation}
  \mathbf{P}_{\ell,h}
  =
  \operatorname{PosAtt}_{h}\bigl(X_{\mathrm{ltt}},\mathbf{F}^{(\ell-1)}\bigr),
  \label{eq:proc_head}
\end{equation}
and $\mathbf{p}_{\ell,h}$ denotes the row of $\mathbf{P}_{\ell,h}$ at the point under consideration. The processor attention stage concatenates that point's feature with these head outputs,
\begin{align}
  \mathbf{h}_{\ell}
  &=
  \bigl[
    \mathbf{f}_{\mathrm{ltt}}^{(\ell-1)}
    \,\|\,
    \mathbf{p}_{\ell,1}
    \,\|\,\cdots\,\|\,
    \mathbf{p}_{\ell,H}
  \bigr],
  \label{eq:proc1}\\
  \bar{\mathbf{h}}_{\ell}
  &=
  \mathrm{InstanceNorm}(\mathbf{h}_{\ell}),
  \label{eq:proc2}\\
  \tilde{\mathbf{h}}_{\ell}
  &=
  \gamma_{\ell}(\mathbf{c})\odot\bar{\mathbf{h}}_{\ell}+\beta_{\ell}(\mathbf{c}),
  \label{eq:proc3}\\
  \mathbf{f}_{\mathrm{ltt}}^{(\ell)}
  &=
  \mathbf{f}_{\mathrm{ltt}}^{(\ell-1)}
  +
  \mathrm{GELU}\bigl(\mathrm{MLP}_{\ell}(\tilde{\mathbf{h}}_{\ell})\bigr).
  \label{eq:proc4}
\end{align}
Equations~\eqref{eq:proc1}--\eqref{eq:proc4} describe the update at a single latent point; instance normalization is computed over the full feature matrix. The concatenation gives $\mathbf{h}_{\ell}$ exactly $D_{\mathrm{p}}=(1+H)D$ channels, so each processor block modulates $q=D_{\mathrm{p}}$ channels. All reported models use $H=1$, and therefore $D_{\mathrm{p}}=2D$. Compared with the corresponding unconditioned processor block,~\eqref{eq:proc3} introduces the condition-dependent modulation. The decoder transfers features from $X_{\mathrm{ltt}}$ to $X_{\mathrm{out}}$ by cross position-attention and, when auxiliary channels $\mathbf{a}$ such as the signed distance to the wall are present, merges them before FiLM. Let $\mathbf{g}_{h}$ be the row, at the query point, of $\operatorname{PosAtt}_{h}(X_{\mathrm{out}},X_{\mathrm{ltt}},\mathbf{F}^{(L)})$. The decoder does not append the query features:
\begin{align}
  \mathbf{g}
  &=
  \bigl[
    \mathbf{g}_{1}
    \,\|\,\cdots\,\|\,
    \mathbf{g}_{H}
  \bigr],
  \label{eq:dec1}\\
  \mathbf{g}
  &\leftarrow
  \mathrm{MLP}_{\mathrm{merge}}\bigl([\mathbf{g}\|\mathrm{MLP}_{\mathrm{aux}}(\mathbf{a})]\bigr),
  \label{eq:dec2}\\
  \mathbf{g}
  &\leftarrow
  \gamma_{\mathrm{dec}}(\mathbf{c})\odot\mathrm{InstanceNorm}(\mathbf{g})+\beta_{\mathrm{dec}}(\mathbf{c}),
  \label{eq:dec3}\\
  \hat{\mathbf{u}}_{\mathrm{out}}
  &=
  \mathrm{MLP}_{\mathrm{de}}(\mathbf{g}).
  \label{eq:dec4}
\end{align}
With one head, the cross-attention output has $D$ channels and the auxiliary merge returns a $D$-channel feature, so $D_{\mathrm{d}}=D$ for every reported model. The auxiliary merge in~\eqref{eq:dec2} is omitted when no auxiliary channels are used; in the analysis below, this corresponds to taking $\mathrm{MLP}_{\mathrm{merge}}$ as the identity. Linear layers use Kaiming initialization~\cite{he2015delving}, and the MLPs use GELU activations. Residual connections in~\eqref{eq:proc4} stabilize deep processors.

\subsection{Preservation of the positional kernel under FiLM}
\label{sec:consistency}
Under the sampling and boundedness assumptions in~\cite{chen2024pit}, discrete position-attention outputs converge in probability to those of a kernel integral operator as the number of sampled points increases. Because FiLM acts pointwise on features, it does not change the distance-based attention weights. To isolate its effect on this convergence, let $\mathcal{K}_{\lambda}^{n}$ denote a discrete position-attention operator on a discretization $X_n$, let $\mathcal{K}_{\lambda}$ denote its limiting operator, and let $\mathcal{M}_{\mathbf{c}}(z)=\gamma(\mathbf{c})\odot z+\beta(\mathbf{c})$ denote the FiLM map~\eqref{eq:film} at a fixed condition $\mathbf{c}$. This $X_n$ is the discretization sequence of the cited result and is distinct from the fixed meshes $X_{\mathrm{ltt}}$ and $X_{\mathrm{out}}$. We compare the two outputs at the same evaluation points, using the Euclidean norm over their feature values. If $\|\gamma(\mathbf{c})\|_{\infty}<\infty$, then

\begin{equation}
  \bigl\|\mathcal{M}_{\mathbf{c}}(\mathcal{K}_{\lambda}^{n}(\mathbf{U}^{n}))
        -\mathcal{M}_{\mathbf{c}}(\mathcal{K}_{\lambda}(\mathbf{U}))\bigr\|
  \le
  \|\gamma(\mathbf{c})\|_{\infty}\,
  \bigl\|\mathcal{K}_{\lambda}^{n}(\mathbf{U}^{n})-\mathcal{K}_{\lambda}(\mathbf{U})\bigr\|.
  \label{eq:composed_convergence}
\end{equation}
Consequently, for a fixed $\mathbf{c}$, bounded FiLM preserves convergence when applied to an already convergent attention output. This observation concerns the affine modulation of an attention output; it does not establish discretization consistency for the complete network, which also contains instance normalization, nonlinear multilayer perceptrons, latent-mesh selection and auxiliary decoder features. Establishing such a result for the complete composition would require additional assumptions on these components.

\subsection{Separation of geometry and condition, and stability in the condition}
\label{sec:theory}
The preceding observation concerns a single modulated attention output. We now examine the dependence of the full network on $\mathbf{c}$, focusing on the \emph{FiLM-only} configuration, in which $\mathbf{c}$ enters only through the FiLM generators and the encoder receives geometric features only. This is the configuration of the reported AirfRANS full-data and BlendedNet++ models. For the analysis we use the matrices of Section~\ref{sec:blocks}: $\mathbf{F}^{(\ell)}\in\mathbb{R}^{N_{\mathrm{ltt}}\times D}$ on $X_{\mathrm{ltt}}$, and $\hat{\mathbf{U}}(\mathbf{c})\in\mathbb{R}^{N_{\mathrm{out}}\times d_{u}}$ collects the predictions $\hat{\mathbf{u}}_{\mathrm{out}}$ at the $N_{\mathrm{out}}$ points of $X_{\mathrm{out}}$. The matrices $\mathbf{H}_{\ell}$, $\tilde{\mathbf{H}}_{\ell}$ and $\mathbf{G}$ are formed likewise from $\mathbf{h}_{\ell}$, $\tilde{\mathbf{h}}_{\ell}$ and $\mathbf{g}$ in~\eqref{eq:proc1}--\eqref{eq:dec4}. Instance normalization in~\eqref{eq:proc2} and~\eqref{eq:dec3} acts on each channel over the points of the set: for $\mathbf{Z}\in\mathbb{R}^{m\times q}$,
\begin{equation}
  \operatorname{InstanceNorm}(\mathbf{Z})_{ik}
  =
  \frac{Z_{ik}-\mu_{k}}{\sqrt{s_{k}^{2}+\epsilon}},
  \qquad
  \mu_{k}=\frac{1}{m}\sum_{i=1}^{m}Z_{ik},
  \qquad
  s_{k}^{2}=\frac{1}{m}\sum_{i=1}^{m}\bigl(Z_{ik}-\mu_{k}\bigr)^{2},
  \label{eq:instnorm}
\end{equation}
with a fixed $\epsilon>0$ ($\epsilon=10^{-5}$ in our implementation).

\begin{proposition}[Separation of geometry and condition]
\label{prop:separation}
Fix a geometry, the input, latent and output point sets, and the auxiliary channels $\mathbf{a}$. In FiLM-only CoPiT, the encoder, processor~\eqref{eq:proc1} and decoder~\eqref{eq:dec1} attention weights, as well as the encoded latent features $\mathbf{F}^{(0)}$ and, when present, the auxiliary embedding $\mathrm{MLP}_{\mathrm{aux}}(\mathbf{a})$, do not depend on $\mathbf{c}$. Consequently,
\begin{equation}
  \hat{\mathbf{U}}(\mathbf{c})
  =
  \mathcal{D}_{\mathbf{c}}\bigl(\mathcal{P}_{\mathbf{c}}(\mathbf{F}^{(0)})\bigr),
  \label{eq:factorization}
\end{equation}
where the processor map $\mathcal{P}_{\mathbf{c}}$ and the decoder map $\mathcal{D}_{\mathbf{c}}$ depend on $\mathbf{c}$ only through $\{(\gamma_{\ell}(\mathbf{c}),\beta_{\ell}(\mathbf{c}))\}_{\ell=1}^{L}$ and $(\gamma_{\mathrm{dec}}(\mathbf{c}),\beta_{\mathrm{dec}}(\mathbf{c}))$, respectively.
\end{proposition}

\begin{proof}
Each attention matrix is determined by pairwise squared distances between fixed point sets, the learned kernel scales and the locality masks defined by distance quantiles. None of these depends on the feature values or $\mathbf{c}$. The encoder uses only these weights and geometric input features, while any auxiliary embedding $\mathrm{MLP}_{\mathrm{aux}}(\mathbf{a})$ depends only on the fixed auxiliary channels. In~\eqref{eq:proc1}--\eqref{eq:dec4}, the condition appears only in the FiLM coefficients.
\end{proof}

\begin{remark}
\label{rem:separation}
Proposition~\ref{prop:separation} has two consequences. First, FiLM does not alter the position-attention weights. When the hypotheses of~\cite[Theorem~2.1]{chen2024pit} hold for an attention layer at a fixed condition,~\eqref{eq:composed_convergence} shows that bounded FiLM preserves the convergence of its output. Second, for a sweep over $K$ conditions on a fixed geometry, the condition-independent quantities can be computed once, giving an idealized cost of $C_{\mathrm{geo}}+K\,C_{\mathrm{cond}}$ rather than $K\,(C_{\mathrm{geo}}+C_{\mathrm{cond}})$, where $C_{\mathrm{geo}}$ and $C_{\mathrm{cond}}$ denote reusable geometry-dependent and condition-dependent computation costs, respectively. When $\mathbf{c}$ is also concatenated with the encoder input, $\mathbf{F}^{(0)}$ becomes condition-dependent, although the attention weights and any condition-independent auxiliary embeddings remain reusable.
\end{remark}

Beyond this separation, we quantify how the prediction varies with $\mathbf{c}$. For a matrix $\mathbf{X}$ with $m$ rows, we use the root-mean-square norm over points, $\nrms{\mathbf{X}}=m^{-1/2}\|\mathbf{X}\|_{F}$. We assume that the conditions lie in a compact set $\mathcal{C}$ and that all FiLM generators satisfy, for $\mathbf{c},\mathbf{c}'\in\mathcal{C}$,
\begin{equation}
  \|\gamma(\mathbf{c})-\gamma(\mathbf{c}')\|_{\infty}\le L_{\gamma}\|\mathbf{c}-\mathbf{c}'\|_{2},
  \qquad
  \|\beta(\mathbf{c})-\beta(\mathbf{c}')\|_{2}\le L_{\beta}\|\mathbf{c}-\mathbf{c}'\|_{2},
  \qquad
  \sup_{\mathbf{c}\in\mathcal{C}}\|\gamma(\mathbf{c})\|_{\infty}\le\Gamma .
  \label{eq:film_lip}
\end{equation}
These bounds hold for multilayer perceptrons with Lipschitz activations on a compact set. In~\eqref{eq:film_lip}, $\gamma$ is the scale applied to the normalized features. On BlendedNet++ the generator emits $\gamma_{\mathrm{raw}}$ and this scale is $\gamma=1+\gamma_{\mathrm{raw}}$, which changes $\Gamma$ but not $L_{\gamma}$; the parameterization is given in Section~\ref{sec:setup}. Let $L_{\mathrm{GELU}}<1.13$ denote a Lipschitz constant of GELU, and let $L_{\mathrm{MLP}}$, $L_{\mathrm{merge}}$ and $L_{\mathrm{de}}$ be Lipschitz constants of $\mathrm{MLP}_{\ell}$ (for all $\ell$), $\mathrm{MLP}_{\mathrm{merge}}$ and $\mathrm{MLP}_{\mathrm{de}}$, applied row by row. The attention maps $\mathbf{F}\mapsto\mathbf{H}_{\ell}$ in~\eqref{eq:proc1} and $\mathbf{F}\mapsto\mathbf{G}$ in~\eqref{eq:dec1} are linear; we denote their Lipschitz constants with respect to $\nrms{\cdot}$ by $\kappa_{\ell}$ and $\kappa_{\mathrm{de}}$. For the attention weight matrices $A_{\ell}^{h}$ and $A_{\mathrm{de}}^{h}$, $h=1,\ldots,H$, valid bounds for the concatenated processor output and the cross-attention output in our implementation are
\begin{equation*}
  \kappa_{\ell}=\Bigl(1+\sum_{h=1}^{H}\|A_{\ell}^{h}\|_{2}^{2}\Bigr)^{1/2},
  \qquad
  \kappa_{\mathrm{de}}=\Bigl(\sum_{h=1}^{H}\|A_{\mathrm{de}}^{h}\|_{2}^{2}\Bigr)^{1/2}\Bigl(\frac{N_{\mathrm{ltt}}}{N_{\mathrm{out}}}\Bigr)^{1/2};
\end{equation*}
without the returned input, the term $1$ in the bound for $\kappa_{\ell}$ is dropped. No value-projection norm appears, because those maps are the identity (Section~\ref{sec:blocks}).

\begin{proposition}[Lipschitz dependence on the condition]
\label{prop:lipschitz}
Under the assumptions above, for FiLM-only CoPiT with fixed point sets and auxiliary channels and for all $\mathbf{c},\mathbf{c}'\in\mathcal{C}$,
\begin{equation}
  \nrms{\hat{\mathbf{U}}(\mathbf{c})-\hat{\mathbf{U}}(\mathbf{c}')}
  \le
  L_{\mathbf{c}}\,\|\mathbf{c}-\mathbf{c}'\|_{2},
  \label{eq:lip_bound}
\end{equation}
with
\begin{equation}
  L_{\mathbf{c}}
  =
  L_{\mathrm{de}}\Bigl(b_{\mathrm{d}}
  +\Gamma\,\epsilon^{-1/2}L_{\mathrm{merge}}\,\kappa_{\mathrm{de}}\,b_{\mathrm{p}}
  \sum_{\ell=1}^{L}\prod_{k=\ell+1}^{L}r_{k}\Bigr),
  \label{eq:lip_const}
\end{equation}
where $r_{k}=1+L_{\mathrm{GELU}}L_{\mathrm{MLP}}\,\Gamma\,\epsilon^{-1/2}\kappa_{k}$, $b_{\mathrm{p}}=L_{\mathrm{GELU}}L_{\mathrm{MLP}}\bigl(\sqrt{D_{\mathrm{p}}}\,L_{\gamma}+L_{\beta}\bigr)$ and $b_{\mathrm{d}}=\sqrt{D_{\mathrm{d}}}\,L_{\gamma}+L_{\beta}$.
\end{proposition}
The proof is given in~\ref{app:proof}. The constant is not sharp; the factor $\epsilon^{-1/2}$ from instance normalization makes it pessimistic. Its structure is nevertheless informative: the condition enters only through the FiLM generators, so every term of $L_{\mathbf{c}}$ contains $L_{\gamma}$ or $L_{\beta}$, while the amplification of condition-induced perturbations is determined jointly by the backbone constants $L_{\mathrm{MLP}}$, $L_{\mathrm{merge}}$, $L_{\mathrm{de}}$, the attention norms, $\Gamma$, $\epsilon$ and the number of blocks. The Lipschitz constants of the generators can be bounded by products of weight norms and activation Lipschitz constants. If $\mathbf{c}$ is also concatenated with the encoder input and the encoder is Lipschitz in $\mathbf{c}$, the same argument extends by propagating the additional initial perturbation $\nrms{\mathbf{F}^{(0)}(\mathbf{c})-\mathbf{F}^{(0)}(\mathbf{c}')}$ through the processor and decoder.
\section{Numerical experiments}
\label{sec:experiments}
We evaluate CoPiT on AirfRANS and BlendedNet++ and compare it with graph-based and transformer-based baselines. Quantities obtained on training data are treated only as optimization diagnostics; all generalization claims are based on held-out data.

\subsection{Experimental setup}
\label{sec:setup}
\subsubsection{Datasets and metrics}

\textbf{AirfRANS}~\cite{bonnet2022airfrans} provides two-dimensional RANS simulations around NACA-family airfoils under freestream states $(U_{\infty},\alpha)$, with unstructured meshes of order $1.8\times 10^{5}$ points. We consider full data (800/200 train/test), scarce data (200/200), Reynolds-number extrapolation and angle-of-attack extrapolation. The full-data model uses four geometric encoder channels (coordinates and unit normals), so $\mathbf{c}=(U_{\infty},\alpha)$ enters only through the FiLM generators. The scarce-data, Reynolds-extrapolation and angle-of-attack models use six encoder channels: the same geometry with $(U_{\infty},\alpha)$ concatenated, while FiLM still reads $\mathbf{c}$. The network predicts $(u_{x},u_{y},p,\nu_{t})$ on the CFD mesh. The decoder additionally uses the signed distance function (SDF) to the wall through a small auxiliary MLP, as in~\eqref{eq:dec2}.

Held-out field errors are computed in the normalized space. The benchmark volume loss is the sum of the pointwise MSEs of $(u_x,u_y,p,\nu_t)$, averaged first over the volume points of each test case and then over test cases. For CoPiT, the squared errors are pooled over all test points; scoring the full-data CoPiT checkpoint with the unmodified benchmark routine gives the same surface loss to four significant digits and a volume loss $1.2\%$ higher, which does not change any comparison in Table~\ref{tab:full}. Because no-slip velocity and turbulent viscosity are prescribed at the wall in this benchmark, the reported surface loss is the corresponding wall-pressure MSE. Force coefficients are obtained by surface integration of pressure and wall shear using the same sampling, gradient and quadrature implementation for every model. We report Spearman correlations of drag and lift, denoted by $\rho_d$ and $\rho_l$; absolute and relative load errors are discussed separately.

\textbf{BlendedNet++}~\cite{sung2025blendednetpp} is a large-scale blended-wing-body (BWB) surface-flow benchmark. We adopt the official forward split with 8\,992 / 1\,000 / 2\,498 train/validation/test cases and predict surface $(C_{p},C_{f,x},C_{f,z})$. Pointwise encoder inputs are six geometric channels (coordinates and unit normals); the 12-dimensional design/flight vector $\mathbf{c}$ drives processor and decoder FiLM only. A hybrid variant that additionally broadcasts $\mathbf{c}$ into the pointwise encoding is evaluated only as an ablation (Section~\ref{sec:blended}).

Targets are $z$-score normalized using training-set statistics only. The test metric is the MSE of each normalized channel over all surface points of a case, averaged over cases, together with the mean over the three channels; the baselines are evaluated on the full surface with the same definition.

\subsubsection{Implementation}

On AirfRANS, the latent mesh has 7\,000 points, four processor blocks, hidden width 256 and one attention head. Encoder and decoder locality are $\tau=0.01$; processor self-attention uses $\tau=1$. All three stages employ the same quantile-masked position-induced attention as PiT. Training minimizes a surface-weighted mean-squared error with Adam~\cite{kingma2015adam}, initial learning rate $10^{-3}$ and cosine annealing~\cite{loshchilov2017cosine} for 398 epochs. The reported AirfRANS scores use the last-epoch checkpoint: the training script writes a single \texttt{model.pth} at the end of the run.

On BlendedNet++, we use hidden width 256, one attention head, four processor blocks and encoder/decoder locality $\tau=0.01$. Pairwise distances are Euclidean squared distances in ambient $\mathbb{R}^{3}$. The latent mesh retains 10\% of the surface points by Z-order sampling. On this dataset, the generator of the reported model and of the hybrid ablation emits $\gamma_{\mathrm{raw}}(\mathbf{c})$, and the scale applied after instance normalization is $\gamma(\mathbf{c})=1+\gamma_{\mathrm{raw}}(\mathbf{c})$, that is, $(1+\gamma_{\mathrm{raw}}(\mathbf{c}))\odot\mathbf{h}+\beta(\mathbf{c})$, so that a vanishing generator output recovers the identity. Because the output layer of the generator carries a bias, this is the same function class as~\eqref{eq:film} with a shifted initialization. AirfRANS applies~\eqref{eq:film} directly, with $\gamma=\gamma_{\mathrm{raw}}$.

Training uses a random subset of 20\,000 surface points per case at every step, and the subset serves as both encoder input and query set. At test time each surface of $n$ points (several tens of thousands) is partitioned into $\lceil n/20\,000\rceil$ random subsets of equal size, at most 20\,000 points each, which are processed in the same way, so that every surface point is predicted. Training uses AdamW with learning rate $5\times 10^{-4}$, weight decay $10^{-5}$, a one-cycle schedule and gradient clipping at unit norm, for 100 epochs. For BlendedNet++, the checkpoint with the lowest validation loss is evaluated once on the test set. Unless stated otherwise, scores are single-run point estimates.

\subsubsection{Baselines}
AirfRANS baselines are MLP~\cite{goodfellow2016deep}, GraphSAGE~\cite{hamilton2017inductive}, PointNet~\cite{qi2017pointnet}, Graph U-Net~\cite{yue2018graph} and Transolver~\cite{wu2024transolver}. On the full-data split we further evaluate matched PiT variants that change only how $(U_{\infty},\alpha)$ is injected: no global condition, input concatenation, processor-only FiLM, decoder-only FiLM, a width-matched concatenation model, and complete CoPiT. BlendedNet++ baselines are reimplemented using the same split, training-set normalization and surface-MSE definition: PointNet, GraphSAGE, Graph U-Net, GNOT~\cite{hao2023gnot}, Transolver and FiLMNet (FiLM without position-induced attention). On BlendedNet++ we additionally compare concatenation-only and hybrid (broadcast globals plus FiLM) variants of the same PiT backbone.

\subsection{AirfRANS results}
\label{sec:AirfRANS}

\Needspace{8\baselineskip}
Tables~\ref{tab:full}--\ref{tab:aoa} report held-out test losses and force-rank correlations for the four AirfRANS splits. All surface and volume entries are evaluation losses rather than optimization losses. The graph-baseline entries are taken from the benchmark evaluation outputs; Transolver and CoPiT predictions are aggregated with the same channel definitions. We first examine the full-data split, followed by scarce-data and extrapolation settings, and then assess the role of conditioning.

\subsubsection{Full-data performance}

\begin{table}[!b]
\caption{AirfRANS full-data test results. The best point estimate in each column is shown in \textbf{bold}. The CoPiT row reports random seed 0; statistics over three seeds are given in the text.}
\label{tab:full}
\centering\footnotesize
\begin{tabular}{@{}lcccc@{}}
\toprule
Model & Surface loss$\downarrow$ & Volume loss$\downarrow$ & $\rho_d\uparrow$ & $\rho_l\uparrow$ \\
\midrule
MLP         & 0.112950 & 0.045625 & $-$0.1169 & 0.9129 \\
GraphSAGE   & 0.066220 & 0.040903 & $-$0.3033 & 0.9651 \\
PointNet    & 0.092511 & 0.112178 & $-$0.0220 & 0.9376 \\
Graph U-Net & 0.038649 & 0.056590 & $-$0.1384 & 0.9667 \\
Transolver  & 0.009120 & 0.037414 & 0.1444 & 0.9981 \\
CoPiT       & \textbf{0.004228} & \textbf{0.028599} & \textbf{0.8388} & \textbf{0.9994} \\
\bottomrule
\end{tabular}
\end{table}

On the full-data split, CoPiT reduces the test surface and volume losses relative to Transolver by approximately $54\%$ and $24\%$, respectively, and has the highest drag and lift rank correlations. Repeating the full-data training with two further random seeds confirms this comparison: over three seeds, CoPiT attains surface and volume losses of $0.00358\pm0.00056$ and $0.0304\pm0.0017$, $\rho_d=0.78\pm0.07$ and $\rho_l=0.99935\pm0.00006$ (mean $\pm$ sample standard deviation), and each of the three CoPiT runs improves on the Transolver run of Table~\ref{tab:full} in all four metrics.

\begin{figure}[t]
\centering
\includegraphics[width=\linewidth,trim=32bp 3bp 6bp 3bp,clip]{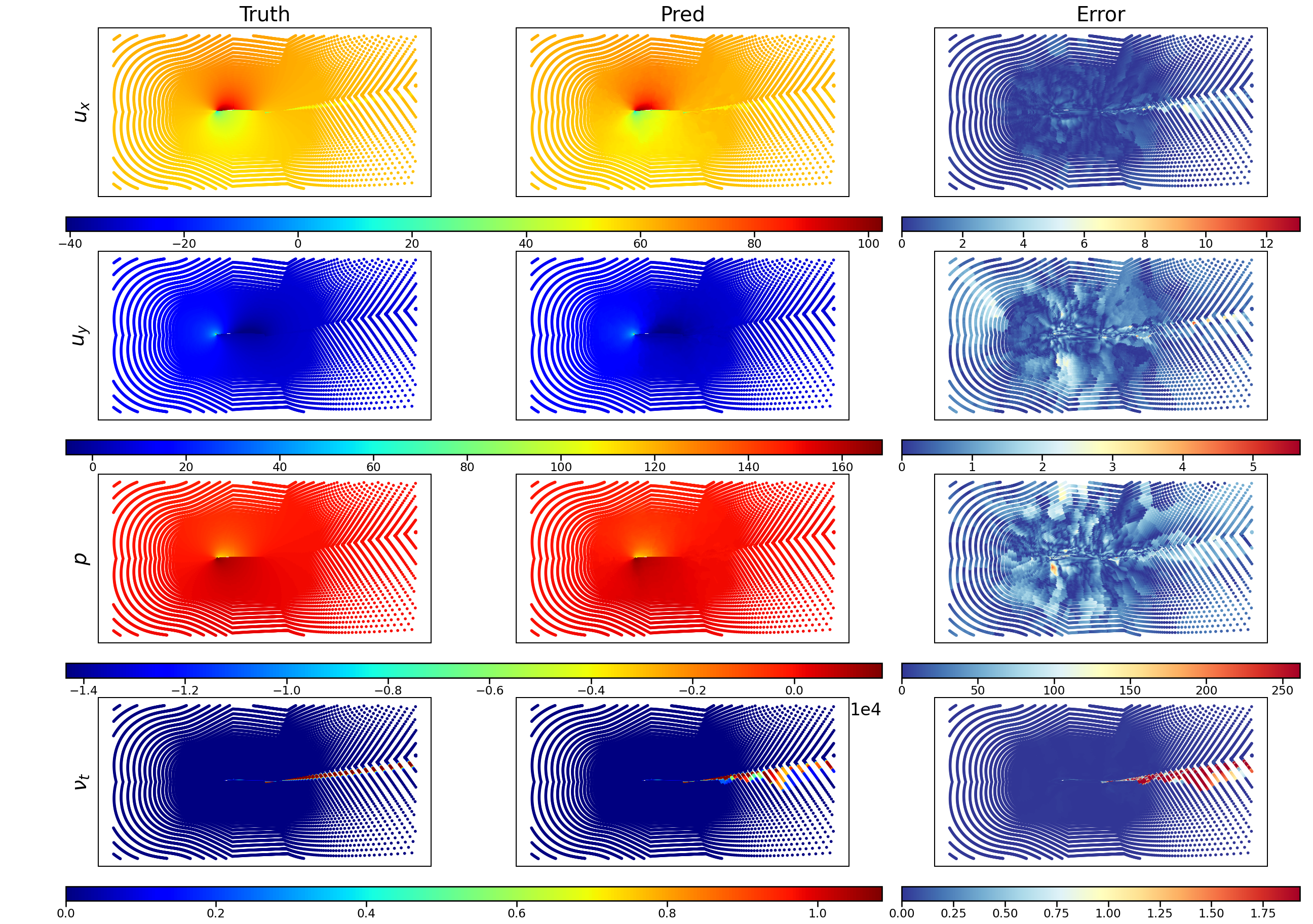}
\caption{AirfRANS full-field maps on one full-data test case ($u_x$, $u_y$, $p$, $\nu_t$): CFD reference, CoPiT prediction and absolute error. Truth and prediction share a color scale per variable; the error uses a separate scale.}
\label{fig:pred_true}
\end{figure}

\begin{figure}[t]
\centering
\includegraphics[width=\linewidth,trim=32bp 3bp 6bp 3bp,clip]{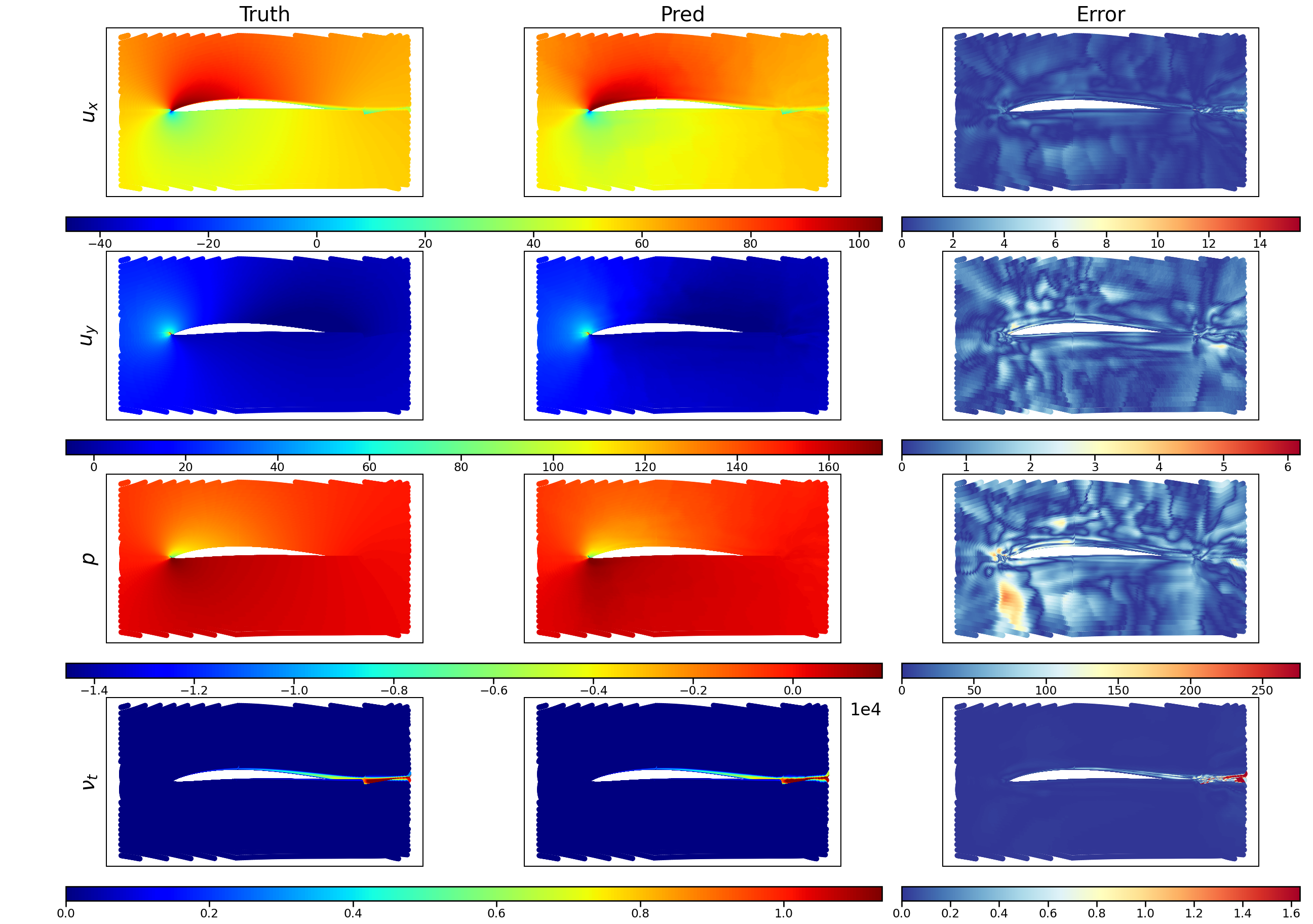}
\caption{Near-wall view of the same AirfRANS test case as Fig.~\ref{fig:pred_true}.}
\label{fig:pred_true_nw}
\end{figure}

Velocity, pressure and eddy viscosity on one full-data test case are shown in Figs.~\ref{fig:pred_true} and~\ref{fig:pred_true_nw}. The predicted fields recover the stagnation region, suction peak, pressure recovery and near-wake, while residuals remain concentrated near the wall. The example is used to visualize spatial error structure rather than to estimate dataset-level accuracy.

\begin{figure}[t]
\centering
\includegraphics[width=\linewidth,keepaspectratio]{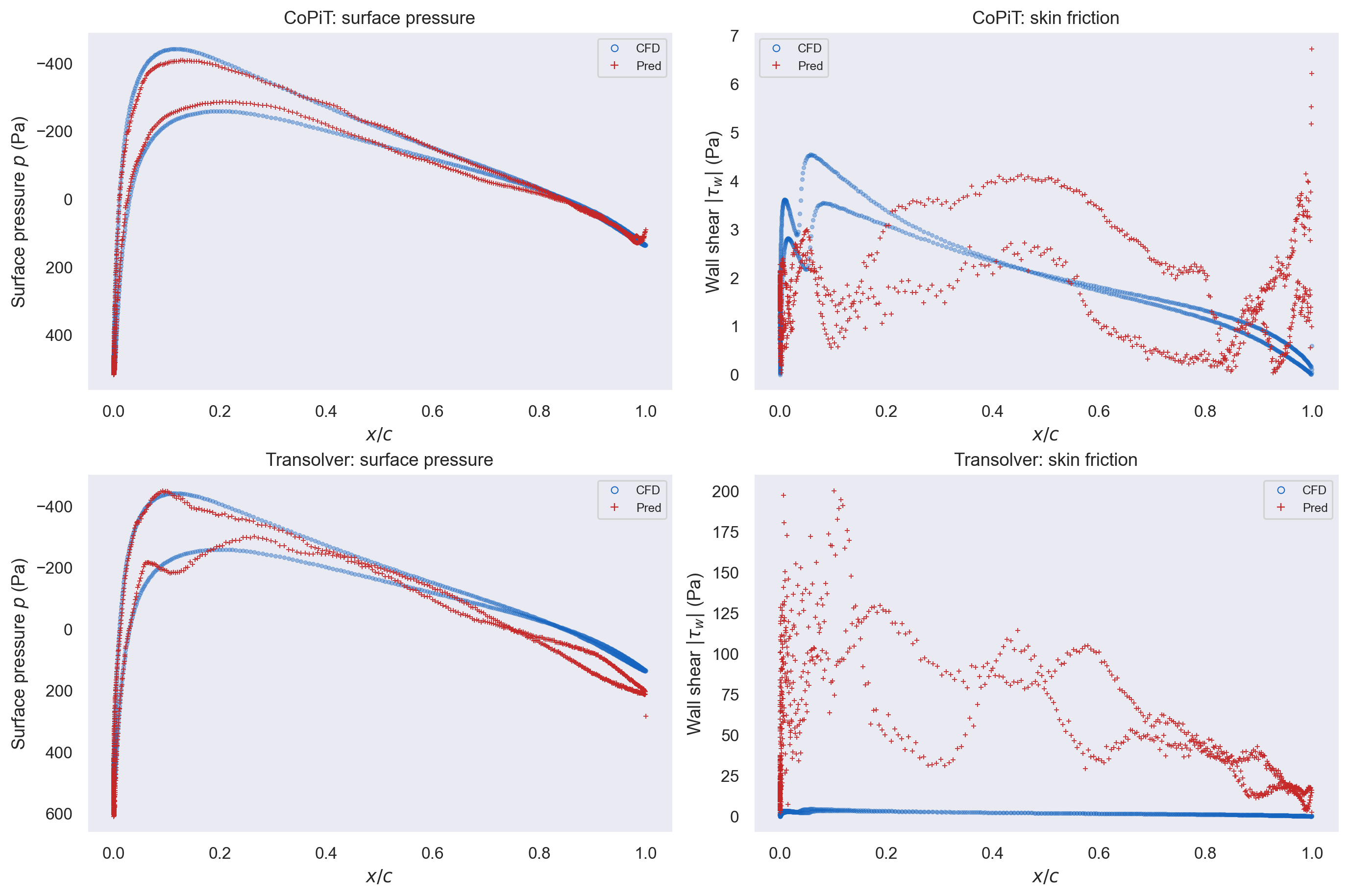}
\caption{Surface pressure and wall shear on one AirfRANS test case (\texttt{airFoil2D\_SST\_31.812\_\ldots}). Top: CoPiT; bottom: Transolver. Blue: CFD; orange: prediction. Wall shear is evaluated from the predicted velocity with a no-slip wall condition.}
\label{fig:AirfRANS_surf_cmp}
\end{figure}

Since $C_{d}$ and $C_{l}$ are obtained by surface integration of pressure and wall shear, Fig.~\ref{fig:AirfRANS_surf_cmp} compares these surface quantities for CoPiT and Transolver on the same test airfoil. Wall shear is computed from the predicted velocity field under a no-slip wall condition. On this sample, the surface-pressure relative $L^{1}$ errors are approximately $0.06$ for CoPiT and $0.18$ for Transolver. The case-wise comparison is descriptive and does not replace test-set aggregate statistics.

\begin{figure}[!b]
\centering
\begin{subfigure}[t]{\linewidth}
  \centering
  \includegraphics[width=0.75\linewidth]{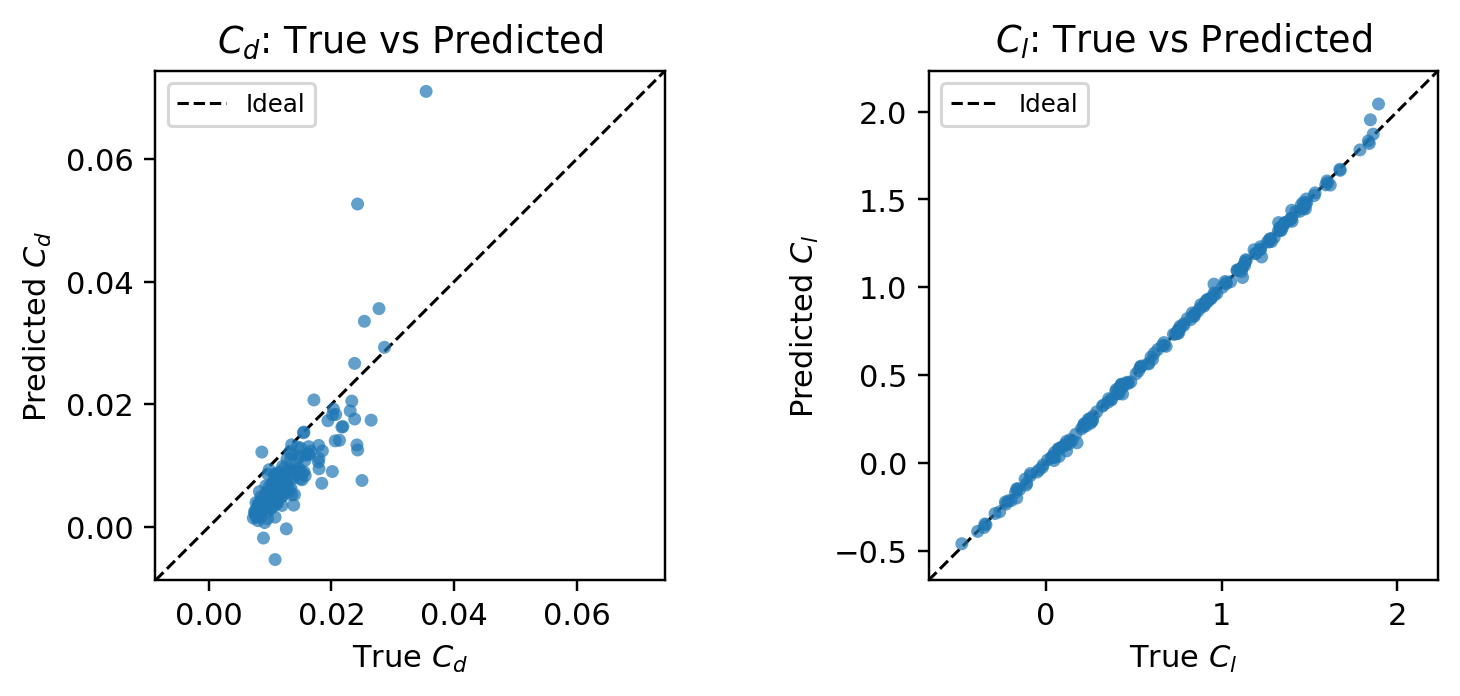}
  \caption{CoPiT}
\end{subfigure}\\[0.3em]
\begin{subfigure}[t]{\linewidth}
  \centering
  \includegraphics[width=0.75\linewidth]{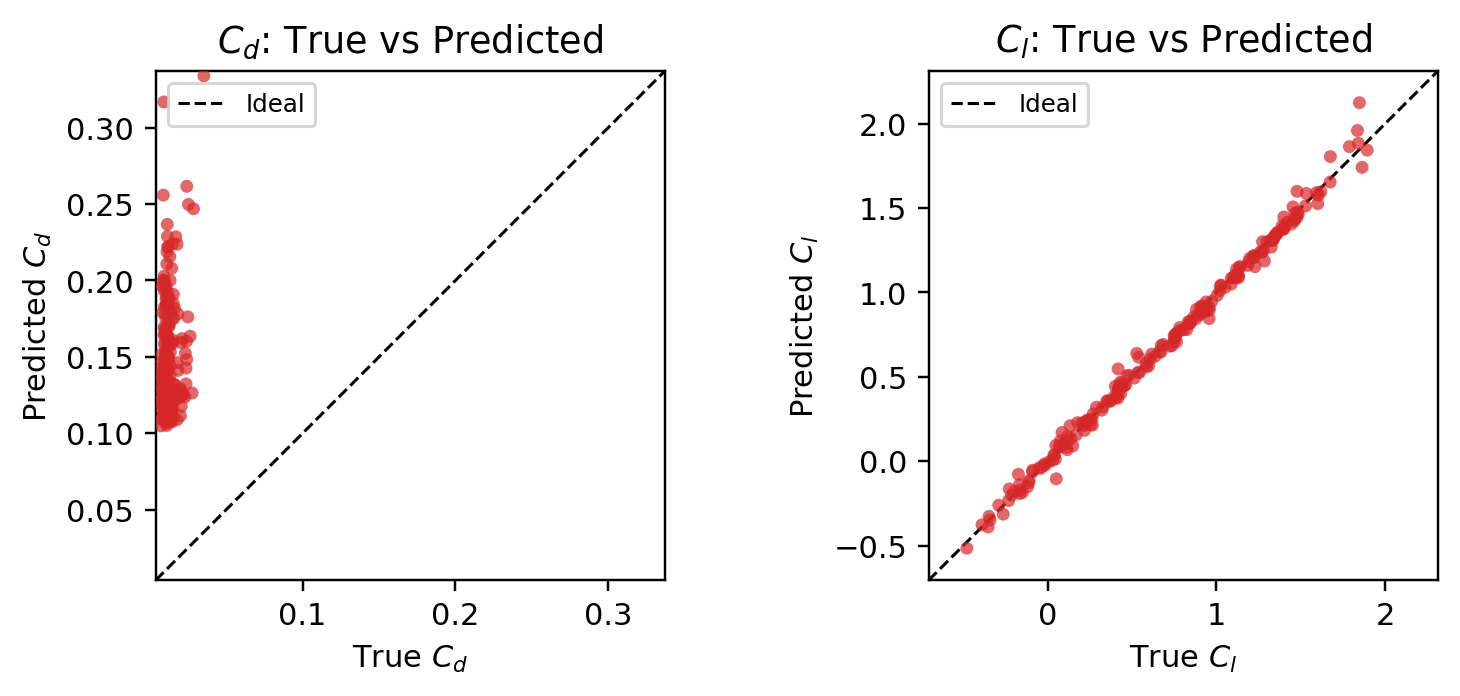}
  \caption{Transolver}
\end{subfigure}
\caption{True versus predicted $C_d$ (left panels) and $C_l$ (right panels) on the AirfRANS full-data test set.}
\label{fig:force_scatter}
\end{figure}

Predicted and reference force coefficients on the 200-case full-data test set are compared in Fig.~\ref{fig:force_scatter} under the same integration protocol. CoPiT yields $\rho_{d}\approx 0.839$ and $\rho_{l}\approx 0.999$, whereas Transolver reaches $\rho_{d}\approx 0.145$ and $\rho_{l}\approx 0.998$: lift ranking is strong for both models, while drag ranking is substantially improved by CoPiT.

For the full-data CoPiT run, the mean relative errors of $C_d$ and $C_l$ are 0.4595 and 0.0467, respectively. Thus the high drag rank correlation is relevant to design screening, but does not imply accurate absolute drag prediction.

\subsubsection{Scarce-data and extrapolation}

\begin{table}[H]
\caption{AirfRANS scarce-data test results (200 training cases).}
\label{tab:scarce}
\centering\footnotesize
\begin{tabular}{@{}lcccc@{}}
\toprule
Model & Surface loss$\downarrow$ & Volume loss$\downarrow$ & $\rho_d\uparrow$ & $\rho_l\uparrow$ \\
\midrule
MLP         & 0.219169 & 0.120089 & $-$0.2423 & 0.9235 \\
GraphSAGE   & 0.194539 & 0.137171 & $-$0.1390 & 0.9813 \\
PointNet    & 0.182676 & 0.147593 & $-$0.0501 & 0.9493 \\
Graph U-Net & 0.147319 & \textbf{0.112956} & $-$0.0950 & 0.9760 \\
Transolver  & 0.086225 & 0.180010 & 0.0673 & 0.9938 \\
CoPiT       & \textbf{0.049273} & 0.237275 & \textbf{0.4231} & \textbf{0.9963} \\
\bottomrule
\end{tabular}
\end{table}

Under scarce data, CoPiT has the lowest surface loss and the highest force-rank correlations, whereas Graph U-Net has the lowest volume loss. The comparatively large CoPiT volume error is driven mainly by turbulent-viscosity prediction and prevents a claim of uniformly superior field accuracy.

\begin{table}[H]
\caption{AirfRANS Reynolds-number extrapolation test results.}
\label{tab:reynolds}
\centering\footnotesize
\begin{tabular}{@{}lcccc@{}}
\toprule
Model & Surface loss$\downarrow$ & Volume loss$\downarrow$ & $\rho_d\uparrow$ & $\rho_l\uparrow$ \\
\midrule
MLP         & 2.089834 & 0.318465 & $-$0.1465 & 0.6423 \\
GraphSAGE   & 0.179743 & 0.318270 & 0.0130 & 0.9271 \\
PointNet    & 0.201310 & 0.332888 & 0.0062 & 0.8985 \\
Graph U-Net & 0.205886 & 0.309486 & 0.0280 & 0.9045 \\
Transolver  & \textbf{0.044703} & 0.332677 & 0.0898 & 0.9477 \\
CoPiT       & 0.045225 & \textbf{0.278162} & \textbf{0.1645} & \textbf{0.9741} \\
\bottomrule
\end{tabular}
\end{table}

For Reynolds-number extrapolation, CoPiT and Transolver have similar surface losses, with Transolver lower by approximately $1.2\%$. CoPiT has the lowest volume loss and the highest rank correlations, although its mean relative drag error is large (approximately $221\%$), which limits the practical interpretation of $\rho_d$.

\begin{table}[H]
\caption{AirfRANS angle-of-attack extrapolation test results.}
\label{tab:aoa}
\centering\footnotesize
\begin{tabular}{@{}lcccc@{}}
\toprule
Model & Surface loss$\downarrow$ & Volume loss$\downarrow$ & $\rho_d\uparrow$ & $\rho_l\uparrow$ \\
\midrule
MLP         & 0.876219 & 0.836325 & 0.0378 & 0.8608 \\
GraphSAGE   & 0.763761 & 0.765209 & 0.0546 & 0.9078 \\
PointNet    & 0.584604 & 0.870350 & 0.1223 & 0.9357 \\
Graph U-Net & 0.696672 & 0.843954 & $-$0.1955 & 0.9344 \\
Transolver  & \textbf{0.237925} & \textbf{0.423247} & 0.0692 & \textbf{0.9957} \\
CoPiT       & 0.261624 & 0.727504 & \textbf{0.5921} & 0.9579 \\
\bottomrule
\end{tabular}
\end{table}

Under angle-of-attack extrapolation, Transolver is more accurate on both field losses and lift ranking, whereas CoPiT has the highest drag rank correlation. CoPiT's mean relative errors of $C_d$ and $C_l$ are approximately $108\%$ and $35\%$, respectively, so the drag-rank result should be interpreted as an ordering metric rather than evidence of accurate extrapolative loads.

\subsubsection{Conditioning ablation}

Before comparing conditioning strategies, we report the model sizes on the full-data split. CoPiT uses about four times fewer trainable parameters than Transolver (Table~\ref{tab:params}).

\begin{table}[!t]
\caption{Trainable parameter counts (AirfRANS full-data models). The six-channel CoPiT checkpoints used for the other splits have 1{,}494{,}538 parameters.}
\label{tab:params}
\centering\footnotesize
\begin{tabular}{@{}lcccccc@{}}
\toprule
 & MLP & GraphSAGE & PointNet & Graph U-Net & Transolver & CoPiT \\
\midrule
Params & 19{,}988 & 29{,}204 & 75{,}244 & 65{,}820 & 5{,}996{,}868 & 1{,}494{,}026 \\
\bottomrule
\end{tabular}
\end{table}

Matched PiT variants are evaluated with the same full-data test protocol as in Table~\ref{tab:full} (Table~\ref{tab:ablate_air}). Removing the freestream condition (geometry-only PiT) collapses both field accuracy and lift ranking. Supplying $(U_{\infty},\alpha)$ by input concatenation, with or without a width increase to match CoPiT's parameter count, recovers lift ranking but remains weaker than FiLM on surface loss, volume loss and $\rho_d$. Decoder-only FiLM attains the lowest surface and volume losses in this family, whereas complete CoPiT (processor and decoder FiLM) attains the highest $\rho_d$. The ablation is reported separately from Table~\ref{tab:full}.

\begin{table}[!t]
\caption{AirfRANS full-data test ablation (same evaluator as Table~\ref{tab:full}). Single-run point estimates; the lowest field loss and highest rank correlation in each column are in \textbf{bold}.}
\label{tab:ablate_air}
\centering\footnotesize
\begin{tabular}{@{}lcccc@{}}
\toprule
Variant & Surface$\downarrow$ & Volume$\downarrow$ & $\rho_d\uparrow$ & $\rho_l\uparrow$ \\
\midrule
Geometry-only PiT & 0.859223 & 1.013652 & 0.3449 & 0.7797 \\
Concatenation only & 0.009882 & 0.035483 & 0.6478 & \textbf{0.9995} \\
Widened concatenation & 0.008969 & 0.037486 & 0.7104 & 0.9993 \\
Processor-only FiLM & 0.007099 & 0.037301 & 0.6774 & 0.9991 \\
Decoder-only FiLM & \textbf{0.002370} & \textbf{0.026978} & 0.8273 & 0.9992 \\
Complete CoPiT & 0.004228 & 0.028599 & \textbf{0.8388} & 0.9994 \\
\bottomrule
\end{tabular}
\end{table}

\FloatBarrier
\subsection{BlendedNet++ results}
\label{sec:blended}

\begin{table}[H]
\caption{BlendedNet++ forward surface regression under a common reimplementation protocol (test set, $N=2{,}498$). Single-run normalized MSE; the lowest point estimate in each column is shown in \textbf{bold}.}
\label{tab:blendednetpp}
\centering\footnotesize
\begin{tabular}{@{}lcccc@{}}
\toprule
Model & Overall$\downarrow$ & $C_p$ & $C_{f,x}$ & $C_{f,z}$ \\
\midrule
\textbf{CoPiT} & \textbf{0.0518} & \textbf{0.0175} & \textbf{0.0978} & \textbf{0.0400} \\
Transolver  & 0.0568 & 0.0188 & 0.1107 & 0.0411 \\
FiLMNet     & 0.0691 & 0.0202 & 0.1356 & 0.0516 \\
PointNet    & 0.0715 & 0.0236 & 0.1336 & 0.0572 \\
GraphSAGE   & 0.0859 & 0.0346 & 0.1416 & 0.0817 \\
Graph U-Net & 0.1327 & 0.1320 & 0.1614 & 0.1046 \\
GNOT        & 0.5234 & 0.5055 & 0.5088 & 0.5559 \\
\bottomrule
\end{tabular}
\end{table}

On the official forward test set, under a common reimplementation protocol, CoPiT attains the lowest overall normalized MSE ($0.0518$) and the lowest error on each of the three channels (Table~\ref{tab:blendednetpp}). Transolver is the closest competitor overall ($0.0568$, about $10\%$ higher). Streamwise skin friction $C_{f,x}$ is the hardest channel for every model. All scores are expressed in the same per-channel $z$-score space used for training and therefore differ in scale from physical relative errors reported elsewhere. The ordering is a single-run point estimate. The GNOT reimplementation completed training (early stopping at validation MSE $0.547$) but yielded substantially higher errors than the other models and is therefore not considered a competitive baseline.

\begin{figure}[H]
\centering
\includegraphics[width=0.98\linewidth]{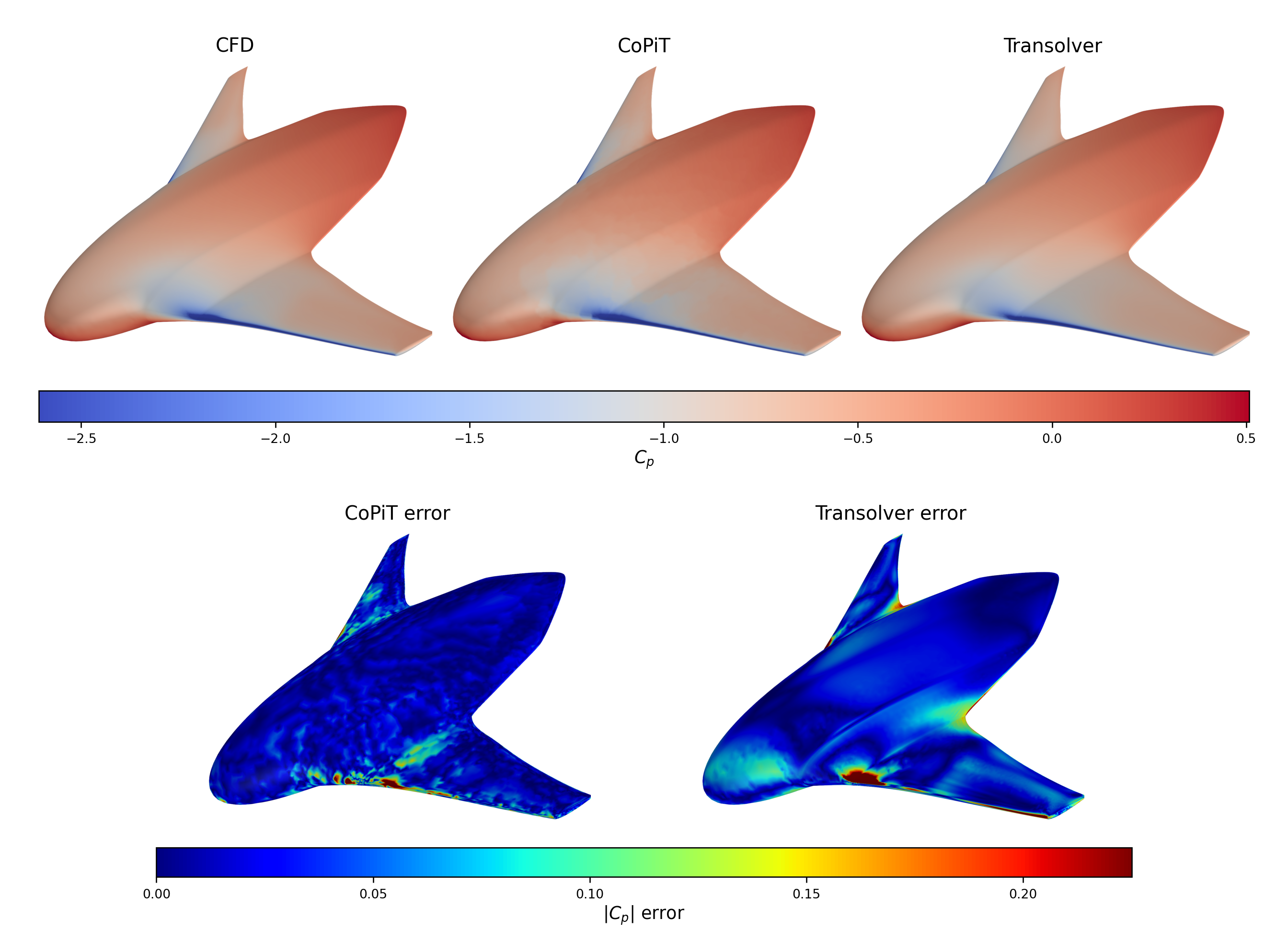}
\caption{BlendedNet++ surface $C_p$ on \texttt{test\_case\_0674}: CFD reference, CoPiT and Transolver predictions (shared color scale), and absolute errors (shared color scale).}
\label{fig:blendednetpp_cp_cases}
\end{figure}

Surface $C_p$ on one test case (\texttt{test\_case\_0674}) is shown in Fig.~\ref{fig:blendednetpp_cp_cases}. Both CoPiT and Transolver recover the large-scale topology; CoPiT exhibits weaker error peaks near the wing--body junction on this sample (relative $L^{1}$ error of order $0.06$ versus $0.12$).

\begin{figure}[H]
\centering
\includegraphics[width=0.98\linewidth]{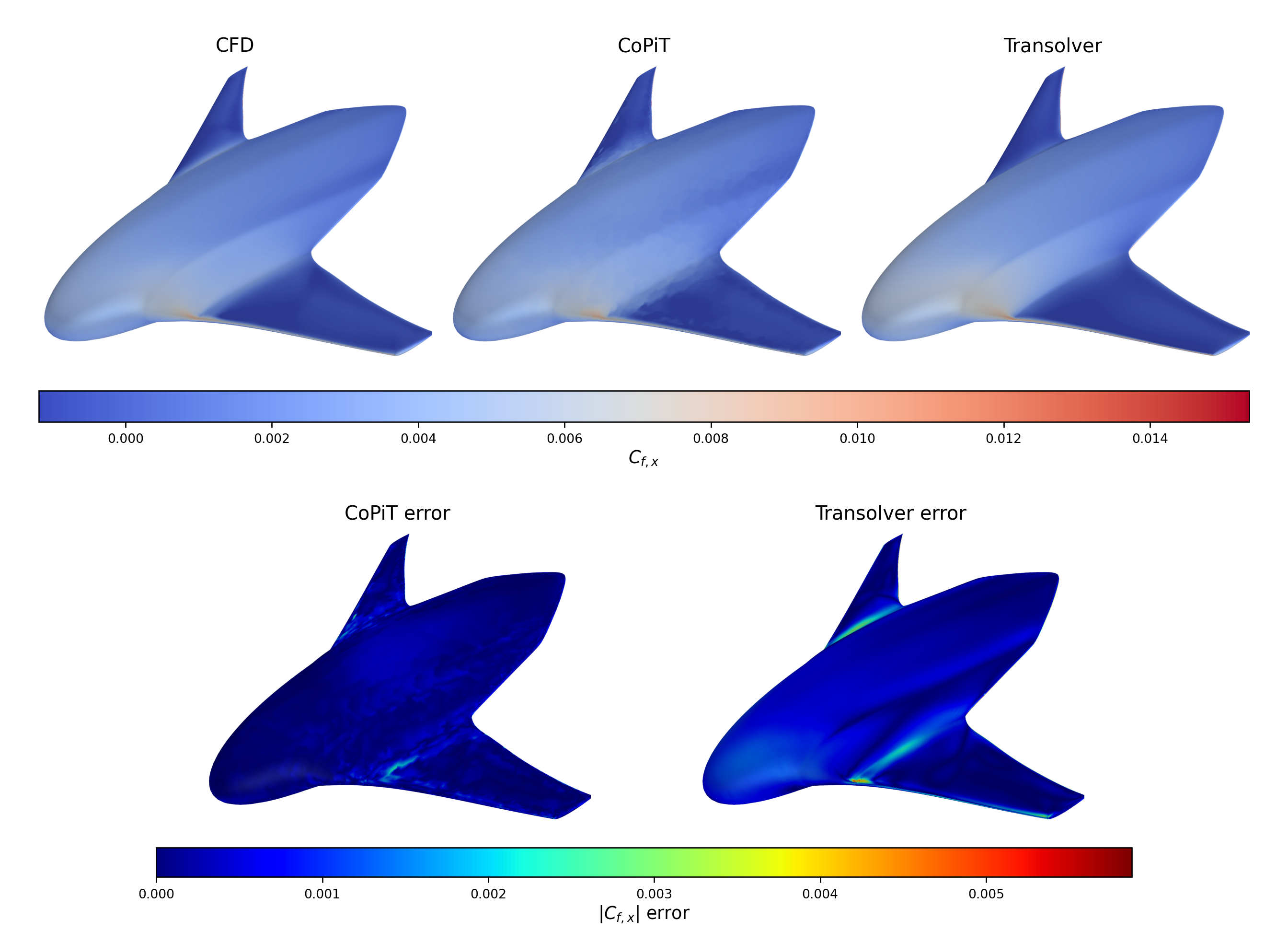}
\caption{BlendedNet++ streamwise skin friction $C_{f,x}$ on the same test case as Fig.~\ref{fig:blendednetpp_cp_cases}: CFD, CoPiT and Transolver (shared color scale), and absolute errors (shared color scale).}
\label{fig:blendednetpp_cfx_cases}
\end{figure}

Streamwise skin friction $C_{f,x}$ on the same case is shown in Fig.~\ref{fig:blendednetpp_cfx_cases}. The relative $L^{1}$ errors are of order $0.11$ for CoPiT and $0.35$ for Transolver; the largest residuals lie near the wing--body junction.

\begin{figure}[H]
\centering
\includegraphics[width=0.98\linewidth]{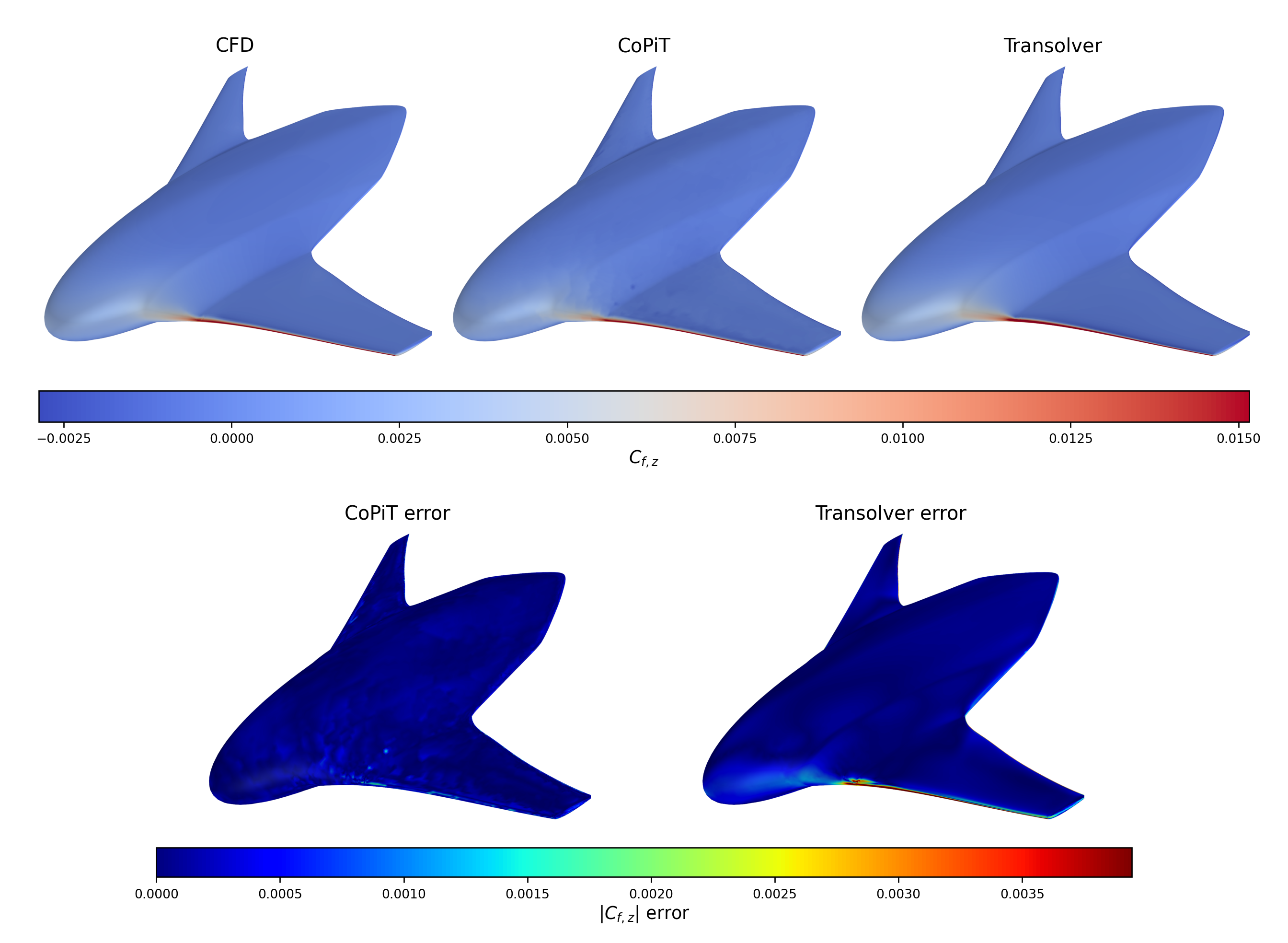}
\caption{BlendedNet++ spanwise skin friction $C_{f,z}$ on the same test case as Fig.~\ref{fig:blendednetpp_cp_cases}: CFD, CoPiT and Transolver (shared color scale), and absolute errors (shared color scale).}
\label{fig:blendednetpp_cfz_cases}
\end{figure}

Spanwise skin friction $C_{f,z}$ is shown in Fig.~\ref{fig:blendednetpp_cfz_cases}. The relative $L^{1}$ errors are of order $0.16$ for CoPiT and $0.46$ for Transolver. These three figures are qualitative and are not used as evidence of population-level superiority.

The conditioning mechanism is isolated on the same PiT backbone, Euclidean metric, official split and 100-epoch budget in Table~\ref{tab:ablate_bnpp}. Concatenation-only (12-dimensional globals broadcast into 18-dimensional point features, no FiLM) yields overall test MSE $0.0741$. Adding FiLM to that encoding (hybrid) reduces the error to $0.0570$. Removing the broadcast and retaining only dual-stage FiLM---the reported CoPiT configuration---gives $0.0530$. Thus, on this split, FiLM is beneficial relative to concatenation, and broadcasting globals into the encoder is not required once FiLM is present. These three runs are single-seed point estimates. Table~\ref{tab:ablate_bnpp} reports, for all three variants, the test metric recorded at the end of training, which uses one random subset of 20\,000 surface points per case; for the reported model this metric is $0.0530$, compared with $0.0518$ on the full surface in Table~\ref{tab:blendednetpp}.

\begin{table}[H]
\caption{BlendedNet++ conditioning ablation on the official forward test set ($N=2{,}498$). Single-run normalized MSE; lowest point estimate in each column in \textbf{bold}. All variants use the same PiT backbone and Euclidean positional metric. Metric on one random 20{,}000-point subset per case, as recorded at the end of training.}
\label{tab:ablate_bnpp}
\centering\footnotesize
\begin{tabular}{@{}lccccc@{}}
\toprule
Variant & Overall$\downarrow$ & $C_p$ & $C_{f,x}$ & $C_{f,z}$ & Params \\
\midrule
Concatenation only & 0.0741 & 0.0263 & 0.1382 & 0.0577 & 1.19M \\
Hybrid (broadcast+FiLM) & 0.0570 & 0.0203 & 0.1066 & 0.0442 & 1.50M \\
FiLM-only (CoPiT) & \textbf{0.0530} & \textbf{0.0181} & \textbf{0.0991} & \textbf{0.0417} & 1.50M \\
\bottomrule
\end{tabular}
\end{table}

\subsection{Computational efficiency}
\label{sec:efficiency}

If $N_{\mathrm{ltt}}$ denotes the latent-mesh size, $L$ the number of processor blocks, $H$ the number of heads and $D$ the hidden width, dense processor self-attention scales as $\mathcal{O}(L N_{\mathrm{ltt}}^{2} H D)$ in arithmetic operations and requires quadratic attention storage unless evaluated in chunks. Encoder and decoder costs depend on the number of retained query--key pairs and scale linearly with the full-mesh size when the number of attended keys per query is bounded. Chunking can reduce peak memory usage without necessarily reducing arithmetic complexity. Wall-clock training times on one NVIDIA A100 (80\,GB) for the main single-run checkpoints are listed in Table~\ref{tab:runtime}. CoPiT is smaller than Transolver on AirfRANS (Table~\ref{tab:params}) and trains in less wall-clock time on that dataset; on BlendedNet++ the two training times are comparable.

\begin{table}[!t]
\caption{Single-run training wall-clock time on one A100. AirfRANS Full uses 398 epochs; BlendedNet++ uses 100 epochs on the official $8{,}992/1{,}000/2{,}498$ split.}
\label{tab:runtime}
\centering\footnotesize
\begin{tabular}{@{}llr@{}}
\toprule
Dataset & Model & Wall-clock \\
\midrule
AirfRANS Full & CoPiT & 11.6\,h \\
AirfRANS Full & Transolver & 19.3\,h \\
BlendedNet++ & CoPiT (FiLM-only) & 25.8\,h \\
BlendedNet++ & Transolver & 21.3\,h \\
BlendedNet++ & Concatenation only & 24.1\,h \\
BlendedNet++ & Hybrid & 25.8\,h \\
\bottomrule
\end{tabular}
\end{table}

For a single operating condition, predicting a full AirfRANS mesh (about $1.8\times10^{5}$ points, chunks of 32\,000) takes $0.075$\,s with CoPiT and $0.098$\,s with Transolver on the same GPU (peak memory $0.9$ and $0.24$\,GB). During inference, CoPiT evaluates the encoder and processor once per field rather than once per query chunk, and evaluates the localized cross-attention on the nearest latent points with the same quantile threshold as in training, aggregated by a sparse matrix product. In single precision, this reproduces the training-time forward pass to a relative difference of $10^{-6}$.

For multi-condition evaluation, the separation of geometry and condition established in Remark~\ref{rem:separation} allows condition-independent computations to be reused. Computing these quantities once per airfoil ($0.29$\,s) and then evaluating 20 angles of attack reproduces the standard forward pass exactly and takes $0.013$\,s per condition, i.e.\ $0.027$\,s per condition including the geometric part.

\subsection{Discussion}
\label{sec:discussion}

\noindent\textbf{Ranking versus absolute error.}\ 
The drag coefficient is a small residual between pressure and friction contributions, so relative error is sensitive to modest absolute deviations and to cases with small reference drag. Spearman correlation measures the ordering of cases and can be informative for design screening, but it does not measure calibration or absolute load accuracy. The full-data run combines high $\rho_d$ with a mean relative drag error of approximately $46\%$; both aspects must therefore be reported.

\noindent\textbf{Role of FiLM.}\ 
On AirfRANS (Table~\ref{tab:ablate_air}), a global condition is necessary: geometry-only PiT is not competitive. FiLM variants improve surface loss and $\rho_d$ relative to concatenation; increasing hidden width without FiLM does not close that gap. Decoder-only FiLM can outperform complete CoPiT on field MSE, so the two FiLM stages are not uniformly additive, while complete CoPiT attains the highest $\rho_d$ in the reported family. On BlendedNet++ (Table~\ref{tab:ablate_bnpp}), FiLM-only CoPiT outperforms both concatenation and hybrid broadcast-plus-FiLM on the same backbone and Euclidean metric.

\section{Conclusion}
\label{sec:conclusion}
We have presented CoPiT, a conditional neural operator for aerodynamic field prediction on irregular meshes. Building on a position-induced Transformer backbone, CoPiT incorporates multi-stage FiLM conditioning to modulate processor and decoder features according to freestream and design parameters. The pointwise modulation leaves position-attention weights unchanged and preserves convergence of an already convergent attention output under bounded modulation. However, establishing discretization consistency for the complete network would require additional analysis.

On the AirfRANS full-data test set, CoPiT achieves the lowest surface and volume losses among the compared published baselines and a higher drag rank correlation than Transolver. Matched ablations show that geometry-only PiT performs substantially worse than conditioned variants, while complete CoPiT improves field losses and drag ranking over concatenation-only alternatives. However, decoder-only FiLM attains lower field losses than complete CoPiT, indicating that the benefits of the two FiLM stages are not uniformly additive. Performance gains also vary across more challenging settings: Graph U-Net achieves the lowest scarce-data volume loss, while Transolver is more accurate on angle-of-attack extrapolation fields and lift ranking. Moreover, the relatively large drag errors show that high rank correlation does not necessarily imply accurate load prediction.

On BlendedNet++, CoPiT achieves the lowest overall normalized test-MSE point estimate among the reimplemented baselines. A matched ablation further favors FiLM-only conditioning over both input concatenation and hybrid broadcast-plus-FiLM, supporting the effectiveness of explicit feature modulation without broadcasting global parameters into the encoder.

Several limitations remain. Proximity-based attention may still couple points that are close in ambient space but separated on the surface. On AirfRANS, prediction errors remain concentrated near the wall. Training also relies entirely on supervised data. Future work includes topology-aware metrics, residual-based regularizers, adaptive latent meshes and tighter coupling of the surrogate to aerodynamic optimization.

\section*{Declaration of competing interest}
The authors declare that they have no known competing financial interests or personal relationships that could have appeared to influence the work reported in this paper.

\section*{Acknowledgements}
\begin{sloppypar}
Z. Qiao's work is partially supported by the NSFC/RGC Joint Research Scheme (grant N\_PolyU5145/24).

Further support for Z. Qiao's work comes from The Hong Kong Polytechnic University projects H-ZGX8 and H-ZGZ1.

This work was partially supported by the National Natural Science Foundation of China (No.~92370108).
\end{sloppypar}

\section*{Data availability}
The AirfRANS dataset is publicly available at \url{https://github.com/Extrality/AirfRANS}. The BlendedNet++ dataset~\cite{sung2025blendednetpp} is publicly available from the Harvard Dataverse, \url{https://doi.org/10.7910/DVN/ICIDK4}.

\section*{Code availability}
The source code and the evaluation records used for the tables will be released upon acceptance. The AirfRANS and BlendedNet++ datasets can be downloaded from their public releases and are not included in the code archive.

\section*{CRediT authorship contribution statement}
\noindent\textbf{Junfeng Chen:} Methodology, Software, Supervision, Writing -- review \& editing;\\
\textbf{Yongchen Fan:} Conceptualization, Methodology, Software, Investigation, Validation, Visualization, Writing -- original draft, Writing -- review \& editing;\\
\textbf{Zhonghua Qiao:} Conceptualization, Formal analysis, Supervision, Funding acquisition, Writing -- review \& editing;\\
\textbf{Kailiang Wu:} Methodology, Formal analysis, Funding acquisition, Supervision, Writing -- review \& editing.

\appendix
\section{Proof of Proposition~\ref{prop:lipschitz}}
\label{app:proof}

Throughout, the point sets are fixed, $\Delta=\|\mathbf{c}-\mathbf{c}'\|_{2}$, and primed quantities are computed with $\mathbf{c}'$. A map applied row by row with Lipschitz constant $L_{\mathrm{row}}$ in the Euclidean norm of a row is $L_{\mathrm{row}}$-Lipschitz with respect to $\nrms{\cdot}$, because $\nrms{\mathbf{X}}^{2}$ is the mean of the squared row norms.

\begin{lemma}
\label{lem:in}
For all $m,q$ and all $\mathbf{X},\mathbf{Y}\in\mathbb{R}^{m\times q}$, the map in~\eqref{eq:instnorm} satisfies
\begin{gather*}
  \nrms{\operatorname{InstanceNorm}(\mathbf{X})-\operatorname{InstanceNorm}(\mathbf{Y})}
  \le\epsilon^{-1/2}\,\nrms{\mathbf{X}-\mathbf{Y}},\\
  \nrms{\operatorname{InstanceNorm}(\mathbf{X})}\le\sqrt{q}.
\end{gather*}
\end{lemma}

\begin{proof}
Instance normalization acts on each column separately, so it suffices to consider the map $f(x)=Px/s(x)$ on $\mathbb{R}^{m}$, where $P=I-\frac{1}{m}\mathbf{1}\mathbf{1}^{\top}$ is the orthogonal projection onto vectors with zero mean and $s(x)=(\|Px\|_{2}^{2}/m+\epsilon)^{1/2}$. Its Jacobian is
\[
  Df(x)=\frac{1}{s(x)}\Bigl(I-\frac{Px\,(Px)^{\top}}{m\,s(x)^{2}}\Bigr)P .
\]
The symmetric matrix in parentheses has the eigenvalues $1$ and $1-\|Px\|_{2}^{2}/(m\,s(x)^{2})\in(0,1]$, and $\|P\|_{2}\le 1$; hence $\|Df(x)\|_{2}\le s(x)^{-1}\le\epsilon^{-1/2}$. By the mean-value inequality on the convex set $\mathbb{R}^{m}$, $f$ is $\epsilon^{-1/2}$-Lipschitz in the Euclidean norm. Summing the squared column differences and dividing by $m$ gives the first claim. For the second, with $\hat{s}^{2}=\|Px\|_{2}^{2}/m$,
\[
  \|f(x)\|_{2}^{2}=\frac{m\,\hat{s}^{2}}{\hat{s}^{2}+\epsilon}\le m ,
\]
so that $\nrms{\operatorname{InstanceNorm}(\mathbf{X})}^{2}\le q$.
\end{proof}

\begin{lemma}
\label{lem:film}
Let $\mathcal{M}_{\mathbf{c}}(\mathbf{Z})=\mathbf{Z}\operatorname{diag}(\gamma(\mathbf{c}))+\mathbf{1}\beta(\mathbf{c})^{\top}$ for $\mathbf{Z}\in\mathbb{R}^{m\times q}$, with generators satisfying~\eqref{eq:film_lip}. If $\nrms{\mathbf{Z}'}\le\sqrt{q}$, then
\[
  \nrms{\mathcal{M}_{\mathbf{c}}(\mathbf{Z})-\mathcal{M}_{\mathbf{c}'}(\mathbf{Z}')}
  \le\Gamma\,\nrms{\mathbf{Z}-\mathbf{Z}'}+\bigl(\sqrt{q}\,L_{\gamma}+L_{\beta}\bigr)\Delta .
\]
\end{lemma}

\begin{proof}
We have
\[
  \mathcal{M}_{\mathbf{c}}(\mathbf{Z})-\mathcal{M}_{\mathbf{c}'}(\mathbf{Z}')
  =(\mathbf{Z}-\mathbf{Z}')\operatorname{diag}(\gamma(\mathbf{c}))
  +\mathbf{Z}'\operatorname{diag}\bigl(\gamma(\mathbf{c})-\gamma(\mathbf{c}')\bigr)
  +\mathbf{1}\bigl(\beta(\mathbf{c})-\beta(\mathbf{c}')\bigr)^{\top}.
\]
Multiplication from the right by a diagonal matrix increases $\nrms{\cdot}$ at most by the largest absolute diagonal entry, and $\nrms{\mathbf{1}b^{\top}}=\|b\|_{2}$. The claim follows from~\eqref{eq:film_lip}.
\end{proof}

\begin{proof}[Proof of Proposition~\ref{prop:lipschitz}]
Let $\delta_{\ell}=\nrms{\mathbf{F}^{(\ell)}-\mathbf{F}'^{(\ell)}}$. By Proposition~\ref{prop:separation}, $\delta_{0}=0$. In block $\ell$, the attention output satisfies $\nrms{\mathbf{H}_{\ell}-\mathbf{H}'_{\ell}}\le \kappa_{\ell}\,\delta_{\ell-1}$, and Lemma~\ref{lem:in} with $q=D_{\mathrm{p}}$ together with Lemma~\ref{lem:film} gives
\[
  \nrms{\tilde{\mathbf{H}}_{\ell}-\tilde{\mathbf{H}}'_{\ell}}
  \le
  \Gamma\,\epsilon^{-1/2}\kappa_{\ell}\,\delta_{\ell-1}
  +\bigl(\sqrt{D_{\mathrm{p}}}\,L_{\gamma}+L_{\beta}\bigr)\Delta .
\]
By~\eqref{eq:proc4},
\[
  \delta_{\ell}
  \le\delta_{\ell-1}+L_{\mathrm{GELU}}L_{\mathrm{MLP}}\,\nrms{\tilde{\mathbf{H}}_{\ell}-\tilde{\mathbf{H}}'_{\ell}}
  \le r_{\ell}\,\delta_{\ell-1}+b_{\mathrm{p}}\Delta ,
\]
and induction from $\delta_{0}=0$ yields
\[
  \delta_{L}\le b_{\mathrm{p}}\,\Delta\sum_{\ell=1}^{L}\prod_{k=\ell+1}^{L}r_{k}.
\]
In the decoder, the auxiliary embedding does not depend on $\mathbf{c}$, so the inputs of $\mathrm{MLP}_{\mathrm{merge}}$ in~\eqref{eq:dec2} differ only in the attention output, and $\nrms{\mathbf{G}-\mathbf{G}'}\le L_{\mathrm{merge}}\,\kappa_{\mathrm{de}}\,\delta_{L}$ after~\eqref{eq:dec2}. Applying Lemma~\ref{lem:in} with $q=D_{\mathrm{d}}$, Lemma~\ref{lem:film} and the Lipschitz constant of $\mathrm{MLP}_{\mathrm{de}}$ to~\eqref{eq:dec3}--\eqref{eq:dec4} yields
\[
  \nrms{\hat{\mathbf{U}}(\mathbf{c})-\hat{\mathbf{U}}(\mathbf{c}')}
  \le
  L_{\mathrm{de}}\bigl(\Gamma\,\epsilon^{-1/2}L_{\mathrm{merge}}\,\kappa_{\mathrm{de}}\,\delta_{L}+b_{\mathrm{d}}\Delta\bigr).
\]
Inserting the bound on $\delta_{L}$ proves~\eqref{eq:lip_bound}--\eqref{eq:lip_const}.
\end{proof}

\FloatBarrier
\bibliographystyle{elsarticle-num}
\bibliography{copit_refs}

\end{document}